\documentclass[a4paper,11pt]{article}
\usepackage[margin = 2.5cm]{geometry}

\usepackage{amsmath}
\usepackage{amssymb}
\usepackage{amsfonts}
\usepackage{amsthm}
\usepackage{mathtools}
\usepackage{bbm}
\usepackage{bm}
\usepackage{slashed}
\usepackage{tensor}
\usepackage{cancel}
\usepackage{braket}
\usepackage{extarrows}
\usepackage{upgreek}
\usepackage{dsdshorthand}
\usepackage[numbers,sort&compress]{natbib}
\usepackage[colorlinks=true
,urlcolor=darkblue
,anchorcolor=darkblue
,citecolor=darkblue
,filecolor=darkblue
,linkcolor=darkblue
,menucolor=darkblue
,linktocpage=true
,pdfproducer=medialab
,pdfa=true
]{hyperref}
\usepackage[capitalize,nameinlink]{cleveref}
\usepackage[labelfont=bf]{caption}

\crefname{figure}{Fig.}{Figs.}
\Crefname{figure}{Fig.}{Figs.}

\crefname{equation}{Eq.}{Eqs.}
\Crefname{equation}{Eq.}{Eqs.}

\crefname{section}{Sec.}{Secs.}
\Crefname{section}{Sec.}{Secs.}
\crefname{appendix}{App.}{Apps.}
\numberwithin{equation}{section}
\allowdisplaybreaks

\usepackage{graphicx}
\usepackage{float}
\usepackage{multirow}
\usepackage{rotating}
\usepackage{dcolumn}
\usepackage{subcaption}
\usepackage{ytableau}
\usepackage{booktabs}
\usepackage{tabularx}
\usepackage{array}

\usepackage{xcolor}
\usepackage{etoolbox}
\usepackage{wasysym}
\usepackage{verbatim}

\usepackage{tikz}
\usepackage{tikz-feynman}

\usetikzlibrary{
    arrows,
    arrows.meta,
    backgrounds,
    calc,
    decorations,
    decorations.markings,
    decorations.pathmorphing,
    decorations.pathreplacing,
    decorations.text,
    fadings,
    intersections,
    shapes,
    shapes.misc,
    tikzmark,
    trees
}

\definecolor{darkgreen}{rgb}{0,0.5,0}
\definecolor{darkblue}{rgb}{0,0,0.6}
\definecolor{purple}{rgb}{0.4,0.2,0.7}
\definecolor{kclred}{RGB}{210,0,30}
\definecolor{kclblue}{RGB}{80,105,180}
\definecolor{lightred}{RGB}{248,220,225}
\definecolor{lightblue}{RGB}{220,230,248}
\definecolor{darkgrey}{RGB}{70,70,70}
\definecolor{coldgrey}{RGB}{78,78,78}
\colorlet{coolgrey}{coldgrey}
\definecolor{hotred}{RGB}{255,58,35}
\definecolor{thermalblue}{RGB}{110,195,235}

\tikzset{
    cross/.style={
        cross out,
        draw=black,
        minimum size=2*(#1-\pgflinewidth),
        inner sep=0pt,
        outer sep=0pt
    },
    branchCut/.style={
        postaction={decorate},
        decoration={
            zigzag,
            segment length=2mm,
            amplitude=2mm
        }
    },
    midarrow/.style={
        postaction={
            decorate,
            decoration={
                markings,
                mark=at position 0.525 with {\arrow{>}}
            }
        }
    }
}

\begin{document}
	
	\thispagestyle{empty}
    
    \begin{flushright}

        \small
    
        KCL-PH-TH/2026-15
    
    \end{flushright}

    \vspace{-1cm}
    
	\begin{center}
		~\vspace{5mm}
		
		\vskip 2cm 
		
		{\LARGE \bf 
			Stochastic Schwinger Effect: de Sitter and beyond
        }
		
		\vspace{0.5in}

		\textbf{Lucas} Vicente García-Consuegra$^1$ and \textbf{Azadeh} Maleknejad$^{2}$

		\vspace{0.5in}
		
        {\footnotesize
		$1$
		{ Department of Physics, King's College London, Strand, London, WC2R 2LS, UK }
		\\
		~
		\\
		$2$
		{ Centre for Quantum Fields and Gravity, Swansea University, Swansea, SA2 8PP, UK}
		\\
        ~
        \\
        ~
        \\
        {\textcolor{darkblue}{\texttt{lucas.vicente\_garcia-consuegra@kcl.ac.uk \quad azadeh.maleknejad@swansea.ac.uk}}}}
\end{center}

	\vspace{0.5in}
	
	\begin{abstract}
    We develop a stochastic formulation of the Schwinger effect in de Sitter spacetime using the Schwinger--Keldysh (in-in) formalism, tailored to particle production by non-stationary gauge backgrounds in the early Universe. Treating the gauge field as a prescribed classical stochastic ensemble, we integrate out massless charged matter and derive the corresponding influence functional and particle-production kernel to leading non-trivial order in the gauge coupling. Conformal invariance allows the result to be extended directly from de Sitter to generic spatially flat FLRW spacetimes, without relying on asymptotic out states. We establish the infrared safety and classicality conditions of the stochastic description and clarify its relation to the conventional static Schwinger effect. We further extend the framework to massless conformally coupled scalars and to weakly coupled non-Abelian gauge sectors, including Standard Model and hidden-sector examples, in regimes where thermal corrections are negligible. Our results provide a general framework for matter creation by stochastic gauge fields in inflation, preheating, and non-thermal BSM sectors, connecting quantum field theory in curved spacetime with early-Universe and high-energy phenomenology.
	\end{abstract}
	
	\vspace{1in}
	
	\pagebreak
	
	\setcounter{tocdepth}{3}
	{\hypersetup{linkcolor=darkblue}\tableofcontents}

\vspace{0.3cm}

\section{Introduction\label{sec:intro}}
Building on our formulation of the stochastic Schwinger effect in four-dimensional flat spacetime~\cite{VicenteGarcia-Consuegra:2025lkh}, we extend this genuinely non-thermal particle-production framework to de Sitter and more general Friedmann-Lemaître-Robertson-Walker (FLRW) spacetimes. We further generalise it to non-Abelian gauge fields in the weak-coupling regime, retaining the leading-order contribution in the gauge coupling. The conventional Schwinger effect provides the natural starting point for this discussion. It describes the non-perturbative production of particle--antiparticle pairs from the quantum vacuum by a sufficiently strong classical background electric field. In the static and homogeneous limit, this process is encoded in the imaginary part of the Euler--Heisenberg effective action \cite{Sauter:1931zz,Heisenberg:1936nmg}.

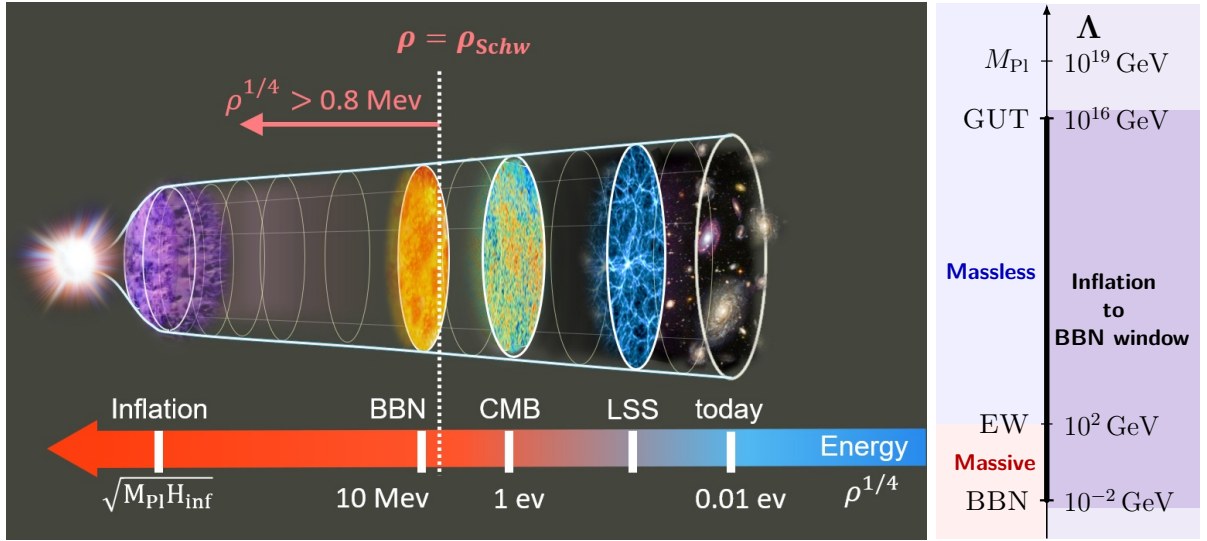
\begin{figure}[ht]
\newsavebox{\cosmichistorybox}
\sbox{\cosmichistorybox}{
    \includegraphics[
        width=0.76\textwidth
    ]{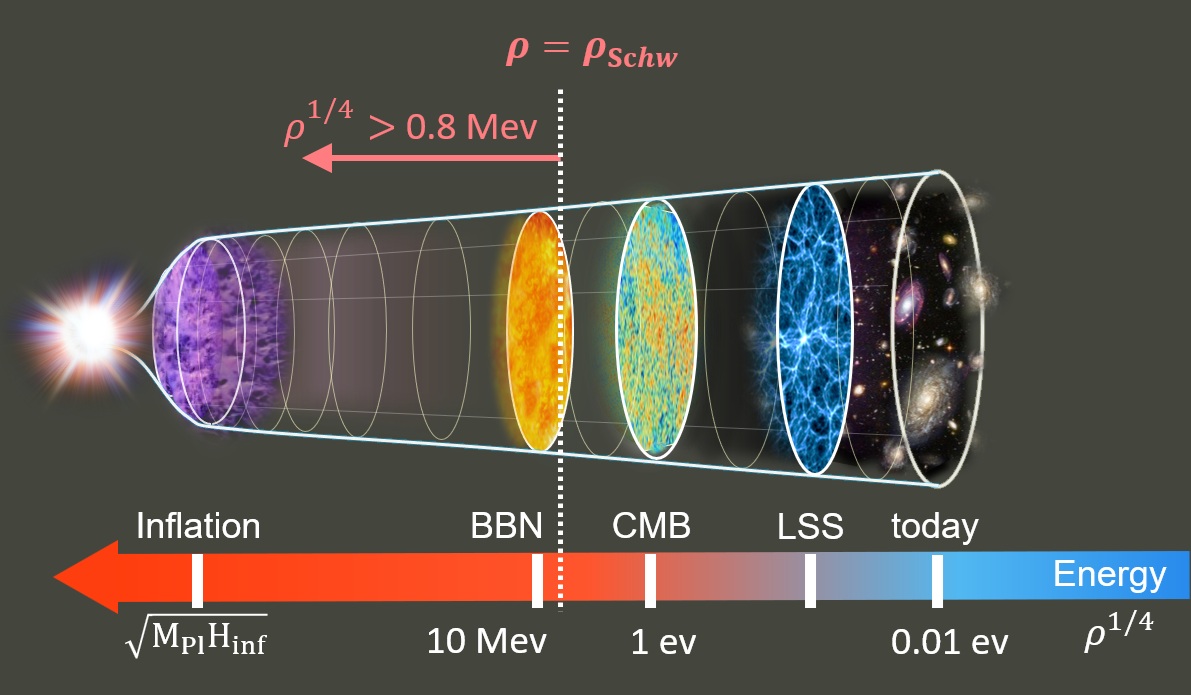}
}
\begin{minipage}[c]{0.76\textwidth}
    \centering
    \usebox{\cosmichistorybox}
\end{minipage}
\hfill
%--------------------------------------------------
% Right-hand TikZ diagram
%--------------------------------------------------
\begin{minipage}[c]{0.23\textwidth}
\centering
\resizebox{!}{
    \dimexpr\ht\cosmichistorybox+\dp\cosmichistorybox\relax
}{
\begin{tikzpicture}[
    >=Latex,
    font=\sffamily,
    every node/.append style={
        scale=1.7,
        transform shape
    },
    tick/.style={
        draw=black,
        line width=0.9pt
    },
    scale label/.style={
        align=right,
        font=\small
    },
    region label/.style={
        align=center,
        font=\sffamily\bfseries\Huge
    }
]
\def\xleft{-2.9}
\def\xright{4}
% Regions
\fill[red!12,opacity=0.5]
    (\xleft,0) rectangle (0,3);

\fill[blue!12,opacity=0.5]
    (\xleft,3) rectangle (0,14);

\fill[purple!20,opacity=0.5]
    (\xright,0) rectangle (0,14);

\fill[purple!35,opacity=0.7]
    (\xright,0.8) rectangle (0,11.2);
    
% Main vertical energy axis
\draw[black,line width=1.15pt,->]
    (0,0) -- (0,14)
    %node[below=0.7pt,font=\sffamily\bfseries\huge]
    %{\qquad\quad $\Lambda$}
    ;
% Ticks
\foreach \y in {1,3,11,12.5}{
    \draw[tick] (-0.15,\y) -- (0.15,\y);
}
% Scale labels
\node[font=\large,anchor=east] at (-0.22,1)
    {\textsc{BBN}};
\node[font=\large,anchor=west] at (0.22,1)
    {$10^{-2}\,\mathrm{GeV}$};
\node[font=\large,anchor=east] at (-0.22,3)
    {\textsc{EW}};
\node[font=\large,anchor=west] at (0.22,3)
    {$10^{2}\,\mathrm{GeV}$};
\node[font=\large,anchor=east] at (-0.22,11)
    {\textsc{GUT}};
\node[font=\large,anchor=west] at (0.22,11)
    {$10^{16}\,\mathrm{GeV}$};
\node[font=\large,anchor=east] at (-0.22,12.5)
    {$M_{\rm Pl}$};
\node[font=\large,anchor=west] at (0.22,12.5)
    {$10^{19}\,\mathrm{GeV}$};
% Region labels
\node[
    region label,
    text=red!70!black,scale=0.4, transform shape
] at (-1.25,2)
    {Massive~};
\node[
    region label,
    text=blue!70!black,scale=0.4, transform shape
] at (-1.25,7)
    {Massless~~};
% Inflationary window
\draw[
    black,
    line width=4pt,
    line cap=round
]
    (0,1) -- (0,11);
% Endpoint markers
\draw[black,line width=1.6pt]
    (-0.18,1) -- (0.18,1);
\draw[black,line width=1.6pt]
    (-0.18,11) -- (0.18,11);
% Vertical labels
\node[
    align=center,
    text=black,
    font=\sffamily\bfseries
,scale=0.9, transform shape] at (1.75,6)
    {Inflation \\to\\~~BBN window};
\node[
    align=center,
    text=black,
    font=\Large
] at (1.1,13.4)
%] at (0.8,13.5)
    %{$\Lambda$};
    {$\boldsymbol{\Lambda}$};
\end{tikzpicture}
}
\end{minipage} 
 \caption{Schematic illustration of the hierarchy of cosmological and particle-physics energy scales. \textbf{Left:} Characteristic energy scale of the Universe, $\rho$, as a function of cosmic history, extending from the inflationary era, characterised by the scale $\sqrt{M_{\rm Pl} H_{\rm inf}}$, to the present-day Universe. The Schwinger energy scale, $\rho_{\rm Schw}$, is shown as a dashed vertical line. \textbf{Right:} The purple shaded region shows the inflationary energy window, bounded above by limits on primordial tensor modes \cite{BICEP:2021xfz} and below by reheating before BBN, relative to the particle-physics scales considered here. Figure adapted from \cite{Maleknejad:2025clz}. }
\label{fig:cosmic-history}
\end{figure}

To place the topic of this work in perspective, we begin with a brief comparison of static Schwinger pair production in four-dimensional Minkowski and de Sitter spacetimes in the presence of a constant electric field. The vacuum-persistence probability and the corresponding pair-production rate are related by
\begin{equation}
    \mathcal{P}=
    \left|\langle 0_{\rm out}|0_{\rm in}\rangle\right|^2
    =
    \exp\left[-\int d^4x\sqrt{-g} \  \Upsilon_{\rm eff}\right],
    \label{eq:vac-persistence}
\end{equation}
where $\Upsilon_{\rm eff}=2\,\text{Im}[\mathcal{L}_{\text{eff}}]$ and $\mathcal{L}_{\text{eff}}$ is the effective Lagrangian density. This formula presumes the existence of well-defined asymptotic in and out states connected by an $\mathcal S$-matrix, and is therefore only strictly available in asymptotically flat spacetimes. Building on the pioneering analyses of Sauter \cite{Sauter:1931zz} and Euler and Heisenberg \cite{Heisenberg:1936nmg}, Schwinger \cite{Schwinger:1951nm} provided the seminal non-perturbative description of vacuum pair production in a constant electric background. Here, for simplicity, we focus on fermionic fields unless stated otherwise and retain only the dominant semi-classical contribution to the imaginary part of the in--out effective action.

\textbf{In flat spacetime:}
For a Dirac fermion of mass $m$ and charge $\mathfrak{q}$ in a constant electric field $E$, the leading Schwinger contribution is
\begin{equation}
    \Upsilon_{\rm eff}^{\rm flat}
    \simeq
    \frac{\mathfrak{q}^2E^2}{4\pi^3}\exp\left(-\frac{\pi m^2}{|\mathfrak{q}E|}\right).
\label{eq:Upsilon-flat}
\end{equation}
For constant parallel electric and magnetic fields, such that $\bm E\cdot\bm B\neq0$, the magnetic field organises the transverse motion into Landau levels and modifies the pair-production rate at leading order \cite{Schwinger:1951nm,Schwartz:2014sze},
\begin{equation}
    \Upsilon^{E\parallel B}_{\rm eff} \simeq \,\frac{\mathfrak{q}^{2}EB}{4\pi^{2}}\,\coth\left(\frac{\pi B}{E}\right)\exp\left(-\frac{\pi m^{2}}{|\mathfrak{q} E|}\right).
\label{eq:Upsilon-fermion-EB}
\end{equation}
A strong-field effective-field-theory framework that systematically incorporates inhomogeneity corrections to the Schwinger effect is developed in \cite{Franchino-Vinas:2025eze}. To illustrate the physical scale involved, consider ordinary quantum electrodynamics (QED), where the electron is the lightest electrically charged particle and therefore sets the lowest threshold for vacuum pair production. The Schwinger critical field, $E_{\text{Schw}} \simeq 1.3\times 10^{18}\,\mathrm{V\,m^{-1}}$, then marks the onset of efficient electron--positron pair production from the vacuum. Expressed in natural units, the corresponding energy scale is
\begin{equation}
    \rho^{1/4}_{\rm Schw} \simeq 0.8\,\mathrm{MeV},
\end{equation}
remarkably close to the electron mass scale. Experimentally, the Schwinger effect has evolved from a long-standing prediction of strong-field QED to analogue realisations in condensed-matter and quantum-simulation platforms, including graphene and cold-atom gauge theories \cite{Schmitt:2023,Zhu:2024}, with further analogue implementations proposed in systems such as superfluid helium \cite{Desrochers:2025}. Yet, field strengths of $\mathcal{O}(10^{18})\mathrm{V\, m^{-1}}$ have never been produced under controlled terrestrial conditions. Strong-field QED experiments are rapidly approaching regimes in which these effects may become experimentally accessible \cite{Fedotov:2023}. Looking ahead, next-generation ultra-intense laser facilities, such as LUXE at DESY \cite{LUXE:2021} and the Extreme Light Infrastructure (ELI)  \cite{Dunne:2008}, aim to approach the strong-field regime in which static Schwinger  pair production from the QED vacuum  may become experimentally accessible.

\textbf{In cosmological spacetimes:} 
Although the Schwinger critical field corresponds to an extreme scale in terrestrial laboratories, it is comparatively modest from a cosmological perspective. As illustrated in \cref{fig:cosmic-history}, the early Universe attained temperatures well above $10\,\mathrm{MeV}$ prior to Big-Bang nucleosynthesis (BBN) \cite{Kolb:1990vq,Weinberg:2008zzc,Baumann:2022mni}, corresponding to energy densities far exceeding the Schwinger scale. It is therefore natural to expect Schwinger-like particle-production processes to have played a role in the early Universe. Since the static Schwinger effect assumes an electric field that is constant in both space and time, the most suitable cosmological setting is cosmic inflation, during which the total energy density remains approximately constant. In our discussion, we take the Bunch--Davies vacuum as the in-vacuum. The definition of an out-vacuum, however, is more subtle, since de Sitter spacetime does not possess a conventional future asymptotic region or a global $\mathcal{S}$-matrix. Nevertheless, a late-time adiabatic out-vacuum can be defined whenever the fermionic modes remain sufficiently oscillatory at late times \cite{Birrell:1982ix,Parker:2009uva}. More precisely, for a constant physical electric field, the late-time frequency behaves as
\begin{equation}
    \omega_{\bk}(\tau)
    \simeq\frac{\mu}{|\tau|},
    \quad\text{where}\quad
    \mu^2
    =\frac{m^2}{H^2}+\frac{\mathfrak{q}^{\,2}E^2}{H^4} - \frac94,
\end{equation}
with adiabaticity parameter $\left|\frac{\omega_{\bk}'}{\omega_{\bk}^2}\right| \simeq \frac{1}{\mu}$. For $\mu\gg1$, the WKB description is valid \cite{Birrell:1982ix,Kobayashi:2014zza}, the positive- and negative-frequency late-time solutions are cleanly separated, and an adiabatic out-vacuum can be introduced. Note that outside this adiabatic regime, the notion of late-time particles, and consequently the in--out interpretation of $\Upsilon_{\rm eff}$, becomes prescription dependent \cite{Kobayashi:2014zza,Hayashinaka:2016dnt}. In this semiclassical regime, and defining the ratio $L=\frac{|\mathfrak{q}E|}{H^2}$, the dominant contribution becomes
\begin{equation}
    \Upsilon_{\rm eff}^{\rm dS}\simeq\frac{H^4}{4\pi^3} \frac{\mu^3}{L} \sum_{s=1}^{\infty} \frac{1}{s^2}\ \exp\left[-2\pi s\left(\mu-L\right)\right],\qquad\mu,L\gg1 .
\label{eq:Gamma-de}
\end{equation}
Unlike the Schwinger effect in flat space \cref{eq:Upsilon-flat}, this de Sitter result is also valid in the massless limit,
\begin{equation}
    \Upsilon_{\rm eff}^{\rm dS}\simeq \frac{\mathfrak{q}^2E^2}{24\pi}\quad\text{for}\quad m=0.
    \label{eq:dS-Upsilon-m0}
\end{equation}
Thus, in flat spacetime pair production is induced solely by the electric field, whereas in de Sitter space the electric and gravitational backgrounds act together \cite{Birrell:1982ix,Parker:2009uva}. For a recent and comprehensive review of particle production induced by the expansion of the Universe, see \cite{Kolb:2023ydq}. In the strong-field regime $ |\mathfrak{q}E| \gg m^2$, massive fermions are produced predominantly through the first term in \cref{eq:Gamma-de},
\begin{equation}
    \Upsilon_{\rm eff}^{\rm dS}\simeq \Upsilon_{\rm eff}^{\rm flat} \quad \text{for} \quad m\neq 0,
\end{equation}
so the de Sitter result reduces to its flat-space counterpart in that regime. This result can be extended to include a background magnetic field with $\bE\cdot\bB\neq0$, as studied in \cite{Bavarsad:2017oyv}. Recent studies have provided useful new perspectives on Schwinger pair production, including a gravitational formulation in conformally flat spacetimes \cite{Franchino-Vinas:2026qhp}, the characterisation of particle--antiparticle entanglement in scalar and spinor QED \cite{Kranas:2025dwz}, and non-local magic in holographic pair production \cite{Grieninger:2026txt}.

\textbf{Axion-Inflation with gauge fields:}
Although Schwinger production can in principle be efficient during inflation, maintaining the required electric field is non-generic, since in the absence of a source the energy density of Maxwell and Yang--Mills fields redshifts as $a^{-4}(\tau)$ and are rapidly diluted by the cosmic expansion. A natural realisation arises in axion inflation models, where the Chern--Simons interaction between the axion and the gauge field can support the gauge field \cite{Anber:2009ua}. While a homogeneous $U(1)$ electric field generally introduces a preferred spatial direction, a non-Abelian gauge field can preserve the symmetries of the cosmological background through an $SU(2)$ subsector \cite{Maleknejad:2011jw, Maleknejad:2011sq,Galtsov:1991un,Dimastrogiovanni:2012ew}, making axion-$SU(2)$ inflation a particularly appealing setting for homogeneous gauge-field backgrounds \cite{Maleknejad:2012fw, Adshead:2012kp,Maleknejad:2016qjz,Dimastrogiovanni:2016fuu}. For comprehensive reviews of gauge fields in inflation, see \cite{Maleknejad:2012fw,Pajer:2013fsa,Komatsu:2022nvu}.

Axion inflation was first recognised in \cite{Lozanov:2018kpk} as a natural and well-motivated setting for the static Schwinger effect in the early Universe, and has since been explored extensively \cite{Mirzagholi:2019jeb,Maleknejad:2019hdr,Domcke:2021fee,vonEckardstein:2024tix,Iarygina:2025ncl}. For non-Abelian gauge fields, the background may also produce gluon pairs through the chromo-Schwinger effect \cite{Maleknejad:2018nxz}. Finally, even in the absence of gauge fields, fermions can be produced directly through a derivative coupling to the pseudoscalar inflaton
\cite{Adshead:2018oaa,Maleknejad:2019hdr}.

\begin{figure}[ht]
    \centering
    \includegraphics[width=0.9\linewidth,trim=7cm 9cm 8cm 9.4cm,clip]{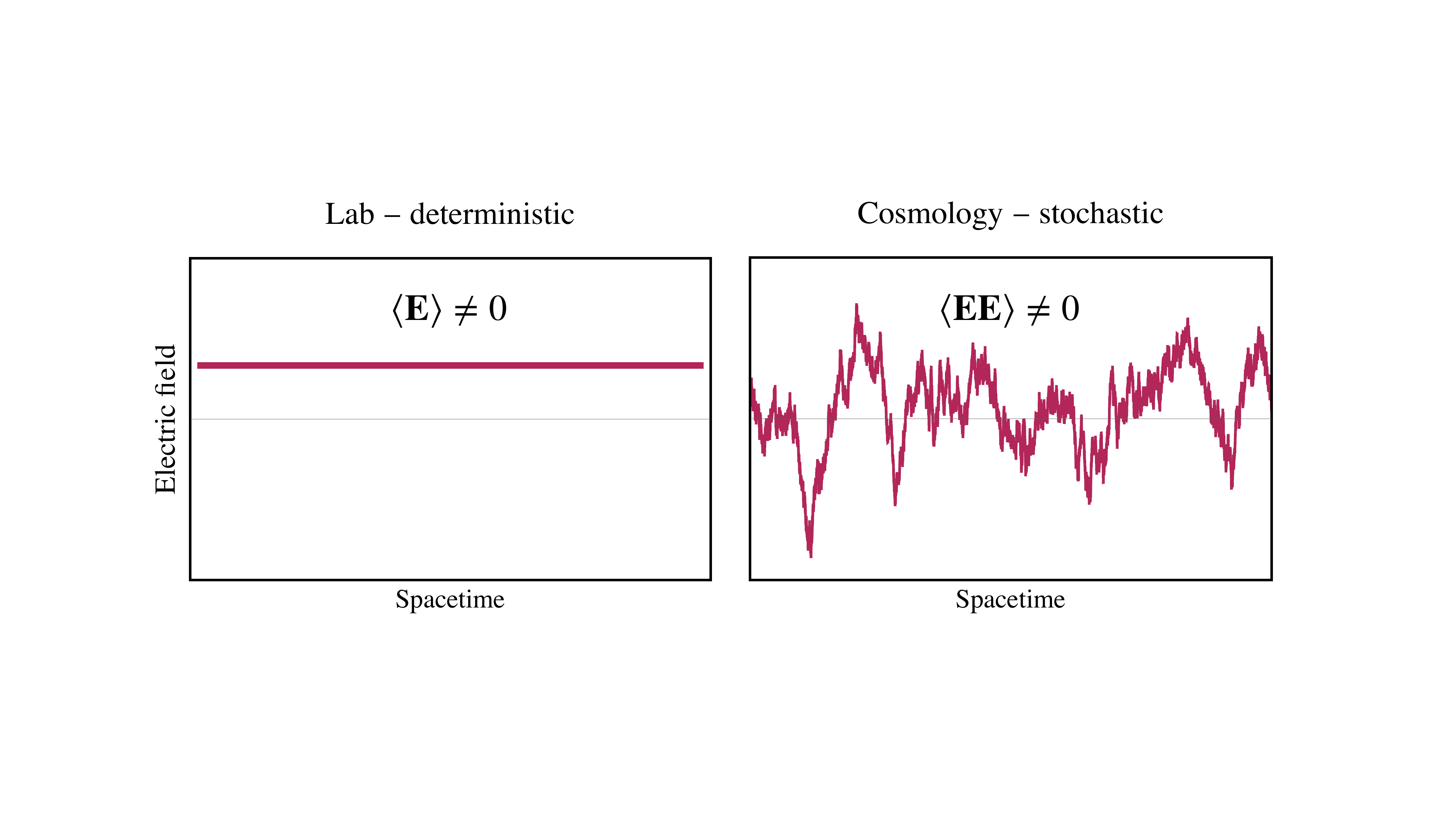}
    \caption{Deterministic vs stochastic electric backgrounds. {\textbf{Left:}}
In conventional laboratory settings, pair production is driven by a coherent electric field with a non-vanishing one-point function, $\langle \bE\rangle\neq0$. {\textbf{Right:}} In cosmological settings, by contrast, the gauge field may be stochastic, with a vanishing mean but non-zero correlations, $\langle \bE\rangle=0$ and $\langle \bE\bE\rangle\neq0$.}
    \label{fig:-s}
\end{figure}

\textbf{Stochastic Schwinger effect:} 
In realistic early-Universe settings, gauge fields are rarely perfectly static or homogeneous. Processes such as axion inflation and the subsequent preheating stage naturally generate stochastic gauge-field backgrounds spanning a broad range of scales (see \cref{fig:-s}), making such backgrounds an unavoidable feature of many early-Universe scenarios \cite{Anber:2009ua,Maleknejad:2011jw,Adshead:2012kp,Adshead:2015pva,Machado:2018nqk}. Lattice simulations of reheating \cite{Adshead:2015pva,Adshead:2017xll,Adshead:2023mvt}, inflation \cite{Caravano:2021bfn,Caravano:2022epk,Figueroa:2023oxc,Sharma:2024nfu,Jamieson:2025ngu,Jamieson:2026fsw}, and audible axions \cite{Machado:2018nqk} further show that these fields can develop a highly stochastic and non-linear character. This motivates the study of the Schwinger effect in stochastic gauge-field backgrounds, as recently undertaken in \cite{VicenteGarcia-Consuegra:2025lkh}. That work was performed in flat spacetime, where time-translation symmetry is preserved. Here, we extend this line of inquiry to systems in which time-translation symmetry is broken, with the expanding Universe and high-energy astrophysical environments involving transient sources as prominent examples. Understanding the interplay between stochastic gauge fields and time-dependent backgrounds is thus essential for a more realistic description of Schwinger pair production in cosmological and astrophysical settings.

\textbf{Non-Thermal Stochastic Particle Production:} The mechanism explored here is part of a broader family of stochastic pair-production phenomena, in which a stochastic background creates particles. Closely related examples include leptogenesis mechanisms driven by the gravitational axial anomaly \cite{Alexander:2004us,Maleknejad:2016dci,Caldwell:2017chz,Alexander:2018fjp,Maleknejad:2024vvf}, baryogenesis during inflation through the $SU(2)$ chiral anomaly \cite{Maleknejad:2020yys,Maleknejad:2020pec}, wash-in leptogenesis \cite{Domcke:2020quw}, and the recently uncovered stochastic production sourced by cosmological perturbations \cite{Maleknejad:2024ybn,Maleknejad:2024hoz,Garani:2024isu,Garani:2025qnm,Redi:2026nzi}.

\subsection*{Summary of results} 

\begin{figure}[ht]
    \centering
    \includegraphics[width=0.9\linewidth,trim=8cm 10cm 10cm 10cm,clip]{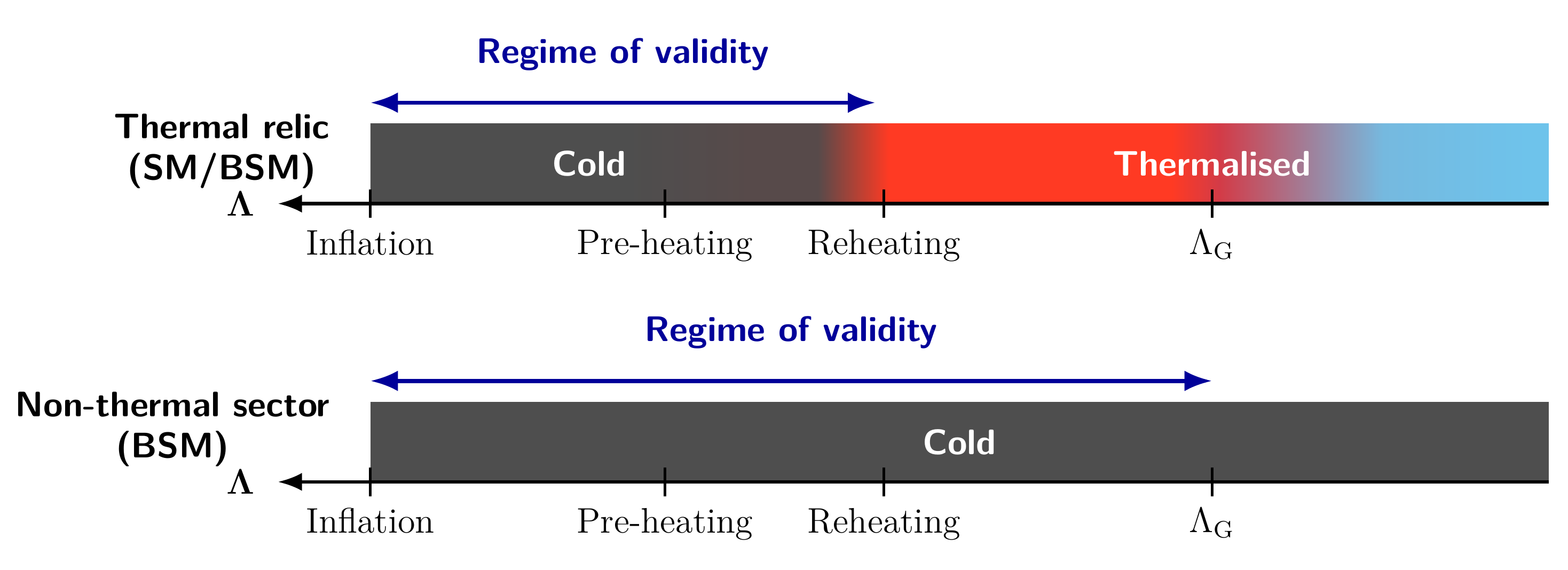}
    \caption{Regimes of validity of the present calculation, in which thermal corrections are neglected. \textbf{Top:} For thermal relic sectors, including Standard Model (SM) or beyond the Standard Model (BSM) degrees of freedom that eventually thermalise with the cosmic plasma, our treatment is most directly applicable during inflation and preheating, while the Universe remains out of thermal equilibrium (see \cref{tab:thermal_history}). After reheating, a thermal plasma is established and thermal corrections generally become important, although they gradually weaken as the Universe expands and cools. \textbf{Bottom:} For non-thermal BSM sectors that never reach equilibrium with the cosmic fluid, $\Gamma_{\rm th}\ll H$, the vacuum calculation can remain applicable over a much broader range of cosmic history. For the non-Abelian extension discussed in \cref{sec:extensions}, one must additionally remain in the perturbative regime, with the characteristic energy scale satisfying $\Lambda\gg\Lambda_{\rm G}$, where $\Lambda_{\rm G}$ denotes the relevant confinement or symmetry-breaking scale. The corresponding domain of validity therefore extends from the high scales of inflation down to the onset of nonperturbative gauge dynamics at $\Lambda\sim\Lambda_{\rm G}$.}
    \label{fig:regime-validity}
\end{figure}

In this work, we develop a stochastic formulation of the Schwinger effect in de Sitter spacetime using the Schwinger--Keldysh (in-in) formalism. By treating the gauge field as a prescribed classical stochastic ensemble and integrating out the quantum matter sector, we derive the influence functional and the corresponding particle-production kernel to leading non-trivial order in the gauge coupling. The imaginary part of the effective action then provides a direct measure of matter creation induced by the stochastic gauge background. Focusing on massless charged particles is motivated both physically and analytically. At the high energy scales relevant to the early Universe, many charged degrees of freedom can be treated as effectively massless (see \cref{fig:cosmic-history}). At the same time, the massless limit allows us to exploit conformal symmetry, which maps the matter sector in conformally flat FLRW spacetimes to its flat-space counterpart and considerably simplifies the analytic treatment. This makes the production kernel more transparent while isolating the effect of the stochastic gauge background itself. Throughout this work, we also neglect thermal corrections, which is well motivated during non-thermal stages such as inflation and the early stages of preheating, or more generally for sectors that remain out of thermal equilibrium with the ambient plasma, $\Gamma_{\rm int}\ll H$ (see \cref{fig:regime-validity} and \cref{tab:thermal_history}).

Our results extend the stochastic Schwinger mechanism from flat spacetime \cite{VicenteGarcia-Consuegra:2025lkh} to expanding cosmological backgrounds and provide a unified framework for particle production by non-stationary gauge fields in the early Universe, where conventional asymptotic in/out states need not exist. This opens a direct connection between quantum field theory in curved spacetime, inflationary particle production, and high-energy phenomenology, with applications ranging from primordial gauge sectors and reheating dynamics to weakly coupled non-Abelian and hidden-sector fields beyond the Standard Model.
Our main results are summarised in \cref{tab:Upsilon-comparison}, where we compare the vacuum-decay rates induced by static and stochastic gauge backgrounds for conformal matter in Minkowski, de Sitter, and spatially flat FLRW spacetimes. We extend the stochastic Schwinger mechanism to weakly coupled non-Abelian gauge sectors relevant at high energies in the early Universe. The result applies when the characteristic scale lies above the corresponding confinement or symmetry-breaking scale and thermal corrections remain negligible, including inflationary/preheating backgrounds and decoupled non-Abelian BSM sectors. Examples of this perturbative extension of the Abelian stochastic pair-production formalism to non-Abelian gauge sectors in the SM and BSM are summarised in \cref{tab:nonabelian-extensions}.

\begin{table*}[ht]
\centering
\renewcommand{\arraystretch}{2.2}
\setlength{\tabcolsep}{6pt}
\resizebox{\textwidth}{!}{
\begin{tabular}{|c|c|c|}
\hline
\large{\textbf{Spacetime}}
& \large{\textbf{Static Schwinger Effect --} 
$\boldsymbol{2\textbf{Im}\Gamma_{\rm eff}^{\rm static}}$}
& \large{\textbf{Stochastic Schwinger Effect --} 
$\boldsymbol{2\textbf{Im}\Gamma_{\rm eff}^{\rm stochastic}}$}
\\
\hline

\large{\textbf{Minkowski}}
&
\rule{0pt}{4.2ex}
\large{\text{Outside the regime of validity}}
\rule[-2.0ex]{0pt}{2.0ex}
&
\rule{0pt}{5.0ex}
$\displaystyle \frac{c_{\rm m} \,\mathfrak q^2}{24\pi} \int_{k^{\mu}} \, (2\pi)^4 \, \Theta(-k_\mu k^\mu)\, \big\langle - F_{\mu\nu}(k) F^{\mu\nu}(-k) \big\rangle_{\rm st} $ \cite{VicenteGarcia-Consuegra:2025lkh}
\rule[-4.0ex]{0pt}{4.0ex}
\\
\hline
\large{\textbf{de Sitter} }
&
\rule{0pt}{5.0ex}
$\displaystyle \frac{c_{\rm m}\mathfrak{q}^2E^2}{24\pi} \,  V_4$ \quad (\textbf{for} \, $|\mathfrak{q}E| \gg H^2$)  \cite{Kobayashi:2014zza}
\rule[-2.5ex]{0pt}{2.5ex}
&
\rule{0pt}{5.0ex}
$\displaystyle \frac{c_{\rm m} \,\mathfrak q^2\pi}{3} \int_{\tau,\tau',\bk}\, \left[ \pi\delta(\Delta\tau) - \frac{\sin(k\Delta\tau)} {\Delta\tau} \right] \big\langle - F_{\mu\nu}(\tau,\bk) F^{\mu\nu}(\tau',-\bk) \big\rangle_{\rm st} $ [\cref{sec:particle_creation}]
\rule[-4.0ex]{0pt}{4.0ex}
\\
\hline

\large{\textbf{ Spatially flat FLRW}}
&
\huge{$\boldsymbol{-}$}
&
\begin{minipage}{0.48\textwidth}
\vspace{1.5ex}
\begin{center} 
\textbf{U(1) or perturbative non-Abelian}
\end{center}
\vspace{1.6ex}

$\displaystyle  \hspace*{-6.45em} \frac{c_{\rm m}\mathfrak{g}^2\pi}{3} \int_{\tau,\tau',\bk} \left[\pi\delta(\Delta\tau)-\frac{\sin(k\Delta\tau)}{\Delta\tau}\right] \left\langle -\Tr_{_{\bR}}\!\left[ F_{\mu\nu}(\tau,\bk) F^{\mu\nu}(\tau',-\bk) \right] \right\rangle_{\rm st}  \text{[\cref{sec:extensions}]}$
\vspace{1.5ex}
\end{minipage}
\\
\hline
\end{tabular}
}
\caption{
For conformal matter fields—namely, a massless Dirac fermion or a massless conformally coupled complex scalar—we compare the effective vacuum-decay rate induced by static and stochastic gauge backgrounds in Minkowski, de Sitter, and FLRW spacetimes. We denote the spin-dependent coefficient by $c_m$, with $c_m=1$ for fermions and $c_m=\frac14$ for scalars.  We define $\Delta\tau=\tau-\tau'$,
where $V_4=\int d^4x\,\sqrt{-g}$ over the field supported region. For compactness, we use the shorthand notation $\int_{x} \equiv \int dx$. }
\label{tab:Upsilon-comparison}
\end{table*}

The remainder of this paper is organised as follows. \cref{sec:QED} introduces massless QED in 4D de Sitter spacetime, emphasising conformal symmetry, the Schwinger--Keldysh formalism, and the stochastic gauge-field backgrounds relevant for cosmology. \cref{sec:particle_creation} develops the stochastic Schwinger effect, deriving the fermionic influence functional, the noise kernel, and then extending the result to spatially flat FLRW backgrounds. \cref{sec:phenomenological_production}  discusses the consistency and limiting regimes of the stochastic description, including infrared safety, classicality of the gauge background, and the relation to the static Schwinger effect. \cref{sec:extensions}  presents further applications and extensions to weakly coupled non-Abelian gauge fields, massless conformally coupled scalars, and non-stationary backgrounds in asymptotically flat spacetimes.
\cref{sec:concl} summarises the main results and outlines future directions. Appendices \ref{app:Dirac}, \ref{app:deets}, and \ref{sec:frequency_resolved_limit} contain mathematical tools and key steps in the derivation of the results.

\medskip
Readers primarily interested in the phenomenological implications may proceed directly to \cref{sec:Noise-Stoch} and the subsequent discussion, while \hyperref[sec:influence_functional]{\namecrefs{sec:Pathintegral}~\ref*{sec:Pathintegral}} through \ref{sec:Q-massless} develop the more formal aspects of the framework.

\subsection*{Conventions and notations:} 
Throughout this work, we work in natural units, $c=\hbar=1$, and use the ``mostly plus" metric signature $(-,+,+,+)$. We also employ \textbf{bold} symbols to denote spatial 3-vectors and the Weinberg convention for Fourier transforms,
\begin{equation*}
    f(\bx)=\int d^3 \bk\  F(\bk)\,e^{i\bk\cdot \bx}
    \quad\text{and}\quad
    F(\bk)=\int \frac{d^3 \bx}{(2\pi)^3}\  f(\bx)\ e^{-i\bk\cdot \bx}.
\end{equation*}
Similarly, we use the shorthand notation for integrals
\begin{equation*}
    \int_{x}\ \equiv \int dx.
\end{equation*}

\section{Massless QED in 4D de Sitter spacetime\label{sec:QED}}
We consider four-dimensional de Sitter spacetime, $dS_{4}$, described in planar coordinates by the line element
\begin{equation}
\label{eq:dS_metric}
    ds^2\equiv g_{\mu\nu}(x)\,dx^\mu dx^\nu= \frac{1}{H^2} \frac{-d\tau^2+d\bx^2}{\tau^2},
\end{equation}
where $a(\tau)=-\frac{1}{H\tau}$ is the scale factor and $x^\mu\equiv(\tau,\bx)$ with $\tau\in\mathbb{R}^{-}$ denoting conformal time and $\bx\in\mathbb{R}^3$ comoving spatial coordinates. Here, $H>0$ denotes the constant Hubble parameter, which sets the curvature scale of the background. In a cosmological setting, \cref{eq:dS_metric} provides the standard approximation to the quasi-de Sitter geometry of an inflationary epoch. Massless quantum electrodynamics on this manifold is described by the action
\begin{equation}
\label{eq:L_QED}
    S_{\text{QED}}
    =
    \int d^{4}x \sqrt{-g}\, \left[-\frac{1}{4}F_{\mu\nu}F^{\mu\nu}+\bar{\Psi}\!\left(i\slashed{\nabla}-\mathfrak{q}\slashed{A}\right)\!\Psi\right],
\end{equation}
where $F_{\mu\nu}\equiv\partial_\mu A_\nu-\partial_\nu A_\mu$ is the field-strength tensor of the Abelian gauge field $A_\mu$ and $\Psi$ is a massless Dirac fermion of charge $\mathfrak{q}=Q\bm{e}$, where $Q$ denotes the charge of the fermion under the $U(1)$ representation and $\bm{e}$ is the gauge coupling. Above, $\nabla_\mu$ denotes covariant differentiation and we use Feynman's notations with $\slashed{\nabla}\equiv\gamma^\mu\nabla_\mu$ and $\slashed{A}\equiv\gamma^\mu A_\mu$ (see \cref{app:Dirac} for a review of spinors in curved spacetimes). As discussed in \cref{sec:intro}, this massless setup is motivated by applications to charged fermions whose masses are negligible at the cosmological scales of interest. These include SM fermions above the electroweak transition, as well as BSM states whose masses are small compared with the characteristic dark-sector scale. Throughout this work, the geometry remains fixed. We therefore neglect gravitational backreaction and quantise only the matter sector on the background spacetime.

Being a gauge theory, the action is invariant under local field redefinitions,
\begin{equation}
\label{eq:local_gauge}
    \Psi(x)\mapsto e^{i\mathfrak{q}\alpha(x)}\Psi(x),\qquad A_\mu(x)\mapsto A_\mu(x)-\partial_\mu\alpha(x),
\end{equation}
where $e^{i\mathfrak{q}\alpha(x)}\in U(1)$ for an arbitrary smooth function $\alpha(x)$. This invariance is a redundancy in the description of the gauge field and must be fixed before performing calculations. For constant $\alpha(x)$, however, \cref{eq:local_gauge} reduces to a global phase rotation of the fermion field. This subset enforces the physical conservation of the electric charge with its associated \emph{vector} Noether current and conservation law given by
\begin{equation}
    j^\mu(x)=\mathfrak{q}\bar\Psi(x)\gamma^\mu(x)\Psi(x),
    \qquad
    \nabla_\mu j^\mu=0.
    \label{eq:J-mu}
\end{equation}
The massless theory further admits an \emph{axial} global $U(1)$ symmetry,
\begin{equation}
    \Psi(x)\mapsto e^{i\beta\gamma^5}\Psi(x), \qquad  A_\mu(x)\mapsto A_\mu(x),
\end{equation}
with $\beta\in\mathbb{R}$ constant and $\gamma^5=\tfrac{i}{4!}\varepsilon_{\mu\nu\rho\sigma}\gamma^\mu(x)\gamma^\nu(x)\gamma^\rho(x)\gamma^\sigma(x)$ denoting the chirality matrix in the Clifford algebra $\text{Cl}_{3,1}(\mathbb{R})$. The corresponding axial current is
\begin{equation}
    j^\mu_5(x)=\bar{\Psi}(x)\gamma^5\gamma^\mu(x)\Psi(x).
\end{equation}
Notably, this current is conserved classically,
\begin{equation}
    \nabla_\mu j^\mu_5=0,
\end{equation}
while at loop level it is violated by the chiral anomaly \cite{Adler:1969gk,Bell:1969ts}.

\subsection{Conformal symmetry}
\label{sec:conf}
A key property of de Sitter spacetime is its conformal flatness. The metric \cref{eq:dS_metric} can be written as
\begin{equation}
    g_{\mu\nu}(x)=a^2(\tau)\,\eta_{\mu\nu},
\end{equation}
where $\eta_{\mu\nu}\equiv\text{diag}(-1,1,1,1)$ is the flat-space metric.\footnote{Note that this is true in every coordinate patch.} At the classical level, the action \cref{eq:L_QED} is invariant under Weyl transformations \cite{Birrell:1982ix},
\begin{equation}
\label{eq:Weyl_rescaling}
    g_{\mu\nu}(x)\mapsto \Omega^2(x)\,g_{\mu\nu}(x),\qquad \Psi(x)\mapsto \Omega^{-3/2}(x)\,\Psi(x), \qquad A_\mu(x)\mapsto A_\mu(x),
\end{equation}
where $\Omega(x)$ is an arbitrary smooth positive function. Choosing $\Omega(x)=a^{-1}(\tau)$ maps the metric to
\begin{equation}
    g_{\mu\nu}(x)\mapsto \eta_{\mu\nu}.
\end{equation}
Now, introducing the conformally rescaled spinor,
\begin{equation}
    \psi(x)\equiv a^{3/2}(\tau)\Psi(x),
    \label{eq:CN-field}
\end{equation}
all dependence on the scale factor cancels and the action becomes
\begin{equation}
    S_{\text{QED}} \mapsto \int d^4x\left[-\frac14F_{\mu\nu}F^{\mu\nu}+\bar{\psi}\!\left(i\slashed{\partial}-\mathfrak{q}\slashed{A}\right)\!\psi\right].
\end{equation}
Similarly, the vector current density takes the standard Minkowski form,
\begin{equation}
    \sqrt{-g}j^\mu(x)\mapsto\mathfrak{q}\bar{\psi}(x)\gamma^\mu\psi(x).
\end{equation}
Classical massless QED on a $dS_4$ background is therefore conformally related to flat-space QED. This observation affects the particle-creation mechanisms available in theory. The rescaled fermion field $\psi$ in the Bunch-Davies vacuum obeys the ordinary flat-space Dirac equation, so its positive-frequency modes are mapped to plane-wave solutions. Consequently, exact de Sitter expansion does not by itself produce massless charged fermions \cite{Parker:2009uva}. Instead, any non-trivial particle production must be induced by external dynamical backgrounds or other ingredients that break the conformal symmetry like a fermion mass \cite{Kolb:2023ydq} or gravitational perturbations \cite{Maleknejad:2024ybn, Maleknejad:2024hoz,Garani:2025qnm,Redi:2026nzi}.

In this work, we focus primarily on de Sitter spacetime, motivated by its central role in inflationary cosmology and its enhanced symmetry. Nevertheless, many of the arguments developed below are not specific to de Sitter. Whenever the analysis relies only on the conformal flatness of the background together with the classical Weyl invariance of the matter sector, the corresponding results extend directly to a generic FLRW spacetime. We will point out this broader applicability where relevant throughout the discussion. For readers primarily interested in the phenomenological applications of our results, we refer directly to \cref{sec:Noise-Stoch} and the subsequent discussion;
\hyperref[sec:influence_functional]{\namecrefs{sec:Pathintegral}~\ref*{sec:Pathintegral}} through \ref{sec:Noise-Stoch} are devoted mainly to the formal development of the framework.

\subsection{Path integral quantisation and the Schwinger--Keldysh formalism}\label{sec:Pathintegral}
As we aim to study \cref{eq:L_QED} at the quantum level, we are interested in computing its partition function. To define the gauge-field path integral, we must remove the redundancy associated with integrating along gauge orbit directions. Formally, we write the configuration space partition function as
\begin{equation}
\label{eq:QED_path}
    \mathcal{Z}=\frac{1}{\mathrm{vol}_{U(1)}} \int \mathcal{D}A_\mu\, \mathcal{D}\bar{\Psi}\,  \mathcal{D}\Psi \  \exp\!\left[iS_{\text{QED}} + i\int d^{4}x\sqrt{-g}\ \left(J^\mu A_\mu+\bar{\zeta}\Psi+\bar{\Psi}\zeta\right)\right],
\end{equation}
where $\mathrm{vol}_{U(1)}$ denotes the volume of the gauge orbit, $J^\mu$ is a classical source for the gauge field, and $\bar{\zeta}$ and $\zeta$ are Grassmann-odd sources for the fermions.  The corresponding Faddeev--Popov--DeWitt determinant can be exponentiated as
\begin{equation}
    \det{}'(-\Box)=\int\mathcal D\bar c\,\mathcal D c\ \exp\big[iS_{\mathrm{ghost}}\big], \qquad  S_{\mathrm{ghost}}  =  -\int d^{4}x\sqrt{-g}\  \bar c\,\Box\, c,
\end{equation}
where $\bar{c}$ and $c$ are Grassmann ghost fields and the prime denotes omission of zero modes of the scalar d'Alembertian, $\Box\equiv \nabla_\mu \nabla^\mu$.  The ghost determinant cancels the gauge orbit volume up to that of the physical homogeneous mode, $\mathrm{vol}_{U(1)_{\mathrm{global}}}$. As the ghost sector is decoupled from the gauge and fermion dynamics, the partition function factorises as
\begin{equation}
\label{eq:Z_QED_flat}
    \mathcal{Z}  = \frac{1}{\mathrm{vol}_{U(1)_{\mathrm{global}}}} \left( \int  \mathcal{D}A_\mu\,  \mathcal{D}\bar{\psi}\,  \mathcal{D}\psi\   e^{iS^\star_{\mathrm{QED}}+i\!\int\! d^4x\, J^\mu A_\mu}   \right)    \times\left( \int \mathcal{D}\bar c\, \mathcal{D}c\  e^{iS_{\mathrm{ghost}}}  \right),
\end{equation}
where, after performing the Weyl rescaling, the gauge-fixed action takes the form
\begin{equation}
\label{eq:QED_gf}
    S^{\star}_{\mathrm{QED}}= S_{\mathrm{QED}}+S_{\xi}
    =
    \int d^{4}x\left[-\frac{1}{4}F_{\mu\nu}F^{\mu\nu}-\frac{1}{2\xi}\left(\partial_\mu A^\mu\right)^2+\bar{\psi}\left(i\slashed{\partial}-\mathfrak{q}\slashed{A}\right)\psi\right].
\end{equation}
Above, the gauge is fixed by imposing the covariant gauge condition with  $\xi\in\mathbb{R}^{+}$ denoting the gauge-fixing parameter \cite{Peskin:1995ev,Schwartz:2014sze}. The residual $U(1)$ volume and the ghost determinant cancel in computations of expectation values and are henceforth omitted. We likewise drop the $\star$ and denote the gauge-fixed action simply by $S_{\mathrm{QED}}$. Finally, since we are ultimately interested in studying the creation of fermions from initial configurations in the fermionic vacuum, we set $\{\bar{\zeta},\zeta\}=0$ from this point on.

\begin{figure}[ht]
    \centering
    \includegraphics[width=\linewidth,trim=11cm 11cm 11cm 11cm,clip]{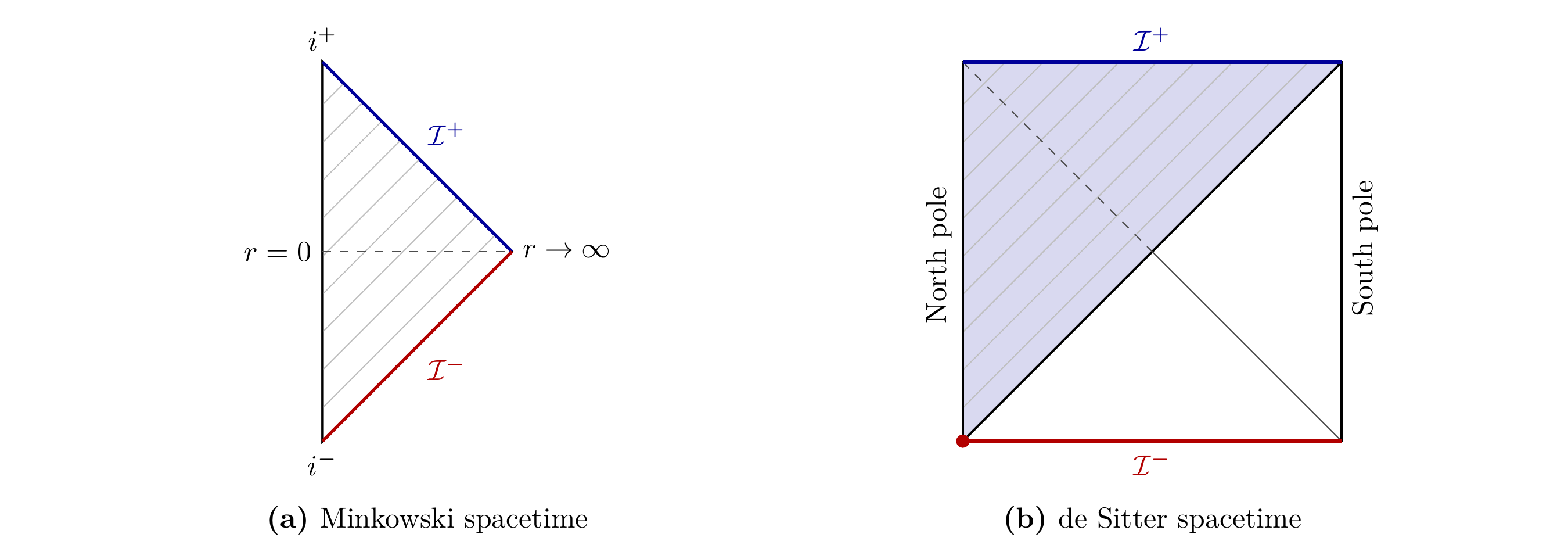}
    \caption{Causal diagrams of Minkowski and de Sitter spacetimes. \textbf{Left panel:} Compactified Minkowski spacetime has null asymptotic boundaries $\mathcal I^\pm$, allowing the definition of in and out states. \textbf{Right panel:} The shaded region is the planar (Poincaré) patch embedded in global de Sitter. It is conformally related to the lower half of the Minkowski diagram with a spacelike future boundary, $\mathcal I^+$, and a null past endpoint, $\mathcal I^-$. Grey lines denote null geodesics like those free fermions and photons follow.}
    \label{fig:PenroseComparison}
\end{figure}

\begin{figure}[ht]
    \centering
    \includegraphics[width=\linewidth,trim=15cm 15cm 15cm 15cm,clip]{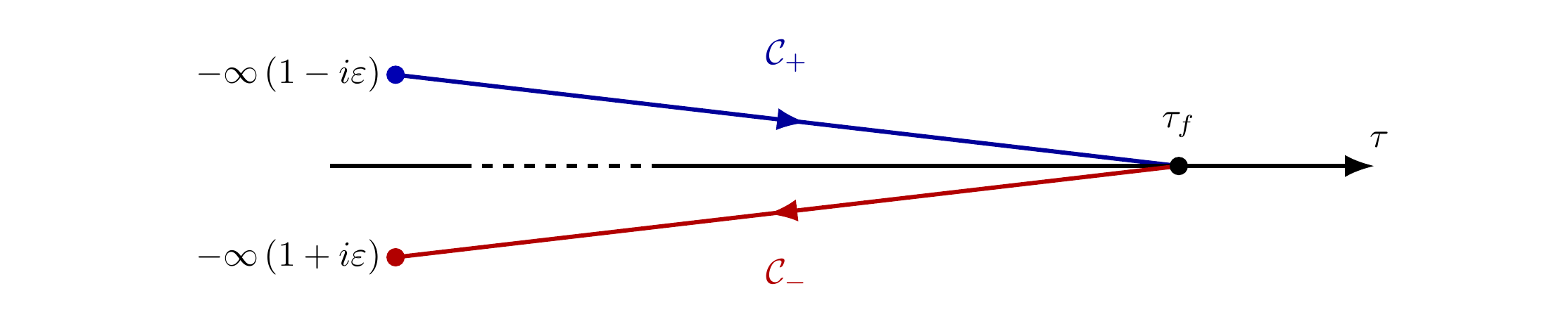}
    \caption{Schwinger--Keldysh closed-time contour. The forward and backward branches, $\mathcal C_+$ and $\mathcal C_-$, are identified at the final time slice $\tau_f$, and the initial state is taken to be the Bunch--Davies vacuum.}
    \label{fig:SKcontour}
\end{figure}

While \cref{eq:Z_QED_flat} formally resembles the standard generating functional of flat-space quantum field theory (QFT), interpreting it as the generator of in-out correlators would require an $\mathcal{S}$-matrix description based on asymptotic particle states. The conformal rescaling \cref{eq:Weyl_rescaling} maps the \emph{local} dynamics of massless QED to their flat-space form, but it leaves the global causal structure of de Sitter spacetime intact. In the planar (Poincaré) patch, conformal time covers only the interval $\tau\in\mathbb R^-$, with future infinity corresponding to the spacelike surface $\tau\to0^-$ rather than null infinity, as shown in \cref{fig:PenroseComparison}. Moreover, de Sitter spacetime does not admit a global timelike Killing vector defining a preferred conserved energy and a unique notion of particles throughout the spacetime. Relevant observables are therefore expectation values evaluated at finite times in a specified quantum state, rather than transition amplitudes between asymptotic particle states. We then formulate the theory using the Schwinger--Keldysh, or in-in, formalism \cite{Schwinger:1960qe,Keldysh:1964ud,Weinberg:2005vy}, which requires doubling the fields and evolving them along a closed time contour, see \cref{fig:SKcontour}.

The full theory closed-time-path (CTP) generating functional is then
\begin{align}
\label{eq:Z_CTP}
    \mathcal{Z}&=\left\langle \widetilde{\mathcal{T}}\exp\left(-i\int d^4x\ J^{\mu\, -} \hat{A}_\mu^-\right) \mathcal{T}\exp\left(i\int d^4x\ J^{\mu\, +} \hat{A}_\mu^+\right) \right\rangle\nonumber\\
    &=\int \mathcal{D}A^{\pm}_\mu\ \rho_{A}[A^+,A^-]\ \int_{|0_{\text{in}}\rangle} \mathcal{D}\bar\psi^{\pm}\,\mathcal{D}\psi^\pm\ \exp\left[i\left(S^+_{\text{QED}}+\int d^4x\ J^{\mu\, +}A_\mu^+\right)\right]\nonumber\\
    &\qquad\times\exp\left[-i\left(S^-_{\text{QED}}+\int d^4x\ J^{\mu\, -}A_\mu^-\right)\right].
\end{align}
Here, $\mathcal{T}$ and $\widetilde{\mathcal{T}}$ denote time ordering and anti-time ordering, while the superscripts $+$ and $-$ label the forward and backward branches of the contour. The density matrix $\hat{\rho}_A$ specifies the initial state of the gauge field, while the fermions are taken to be in the Bunch--Davies vacuum, $|0_{\text{in}}\rangle$. At the final time $\tau_f$, the field configurations on the two branches are identified and traced over. This closes the contour and ensures the the theory remains unitary.

\subsection{Stochastic gauge-field backgrounds in cosmology}\label{sec:stoch-BG}
Gauge fields generated during inflation or preheating are not generally described by a coherent configuration. Their evolution depends on fluctuating initial data and is more naturally described by a stochastic ensemble of classical fields. For each realisation, we then expand the partition function \cref{eq:Z_CTP} about the classical saddle
\begin{equation}
    (A_\mu,\psi)=(A_\mu^{\text{st}},0),
\end{equation}
where $A_\mu^{\text{st}}$ is a stochastic realisation of the gauge field, while fermion fluctuations remain quantum mechanical. This generates the fermion-loop corrections to the classical Maxwell action. Stochastic fields are then treated as random c-numbers with stochasticity encoded in its initial data, $\hat{\rho}_A$. We denote the corresponding ensemble average by $\langle\cdots\rangle_{\mathrm{st}}$ and assume a vanishing one point function and non-zero variance,
\begin{equation}
\label{eq:stoc}
    \langle {A}_\mu(x)\rangle_{\text{st}} = 0,\qquad\qquad\langle {A}_\mu(x) {A}_\nu(y)\rangle_{\text{st}}=\mathcal{G}_{\mu\nu}(x,y).
\end{equation}
Microscopically, such an ensemble may arise as the classical limit of a quantum gauge field state. Here, however, its correlation functions are treated as external input. This type of description arises in several early universe settings, including inflation \cite{Anber:2009ua,Maleknejad:2011jw,Adshead:2012kp,Maleknejad:2012fw}, preheating \cite{Adshead:2015pva}, and other non-equilibrium processes \cite{Machado:2018nqk,Weinberg:2008zzc}.

For a statistically homogeneous ensemble, it is convenient to expand each realisation in spatial Fourier modes, 
\begin{equation}
    {A}_\mu(\tau,\bx)=\int {d^3\bk}\  {A}_\mu(\tau,\bk) \, e^{i\bk\cdot\bx}+\text{c.c.},
\end{equation}
where $\text{c.c.}$ denotes  complex conjugation with the coefficients are given in the helicity basis as
\begin{equation}
    {A}_\mu(\tau,\bk) = \sum_{\sigma=\pm} \epsilon_\mu^{(\sigma)}(\hat{\bk})\left[{\alpha}_{\sigma,\bk}\, A_\sigma(\tau,k)+{\alpha}_{\sigma,-\bk}^*\, A^*_\sigma(\tau,k)\right].
\end{equation}
Here, $\hat{\bk}\equiv \bk/k$, $\sigma=\pm$ labels helicity, and $\epsilon_\mu^{(\sigma)}(\hat{\bk})$ are polarisation vectors transverse to the ${\bk}$-direction. The deterministic functions $A_\sigma(\tau,k)$ describe the time evolution, whereas the random coefficients $\alpha_{\sigma,\bk}$ encode realisation-dependent data. We choose their normalisation so that
\begin{equation}
    \langle{\alpha}_{\sigma,\bk}\,{\alpha}_{\sigma',\bk'}^*\rangle_{st}
    =\delta_{\sigma\sigma'}
    \delta^{(3)}(\bk-\bk').
\end{equation}
With this convention, the spectral information is carried by the mode functions. Equivalently,
\begin{equation}
    \langle {A}_\mu(\tau,\bk) {A}^*_\nu(\tau',\bk')\rangle_{st}
    =
    \delta^{(3)}(\bk-\bk')\,\mathcal{G}_{\mu\nu}(\tau,\tau';k).
\end{equation}
Although it is convenient to define the ensemble using $A_\mu$, physical results must remain gauge invariant. The stochastic input can therefore be specified in terms of the two-point function of the field-strength $F_{\mu\nu}$, or equivalently by quadratic electric and/or magnetic field correlators.

\section{Stochastic Schwinger effect in de Sitter and beyond}
\label{sec:particle_creation}

In asymptotically flat spacetimes, the imaginary part of the in–out effective action directly measures the instability induced by the background, through the vacuum-decay probability
\begin{equation}
    \mathcal P_{\text{decay}}
    =
    1-
    \bigl|\langle 0_{\rm out}|0_{\rm in}\rangle_A\bigr|^2
    \simeq
    2\,\text{Im}\Gamma_{\text{eff}}=\int d^4x\sqrt{-g} \ \Upsilon_\text{eff}(x),
\end{equation}
where the last relation holds in the semi-classical regime, $\text{Im}\Gamma_{\text{eff}}\ll1$. A positive imaginary part of the effective action therefore serves as the field-theoretic signature of pair production. The paradigmatic realisation of this mechanism is the static Schwinger effect \cite{Schwinger:1951nm} reviewed in \cref{sec:intro}. In our previous work \cite{VicenteGarcia-Consuegra:2025lkh}, we generalised this picture to fluctuating electromagnetic backgrounds, introducing the \emph{stochastic Schwinger effect}: pair creation driven not by a coherent classical field but by stochastic fluctuations characterised by their statistical correlations, as in \cref{fig:-s}.

This construction relies on the existence of well-defined asymptotic in and out vacua, connected by an $\mathcal{S}$-matrix, so that the ${\rm Im}\,\Gamma_{\text{eff}}$ can be interpreted as governing a transition probability between them. De Sitter space admits no such structure: the absence of a global timelike Killing vector removes any preferred notion of asymptotic particle states, and with it the $\mathcal{S}$-matrix on which the persistence amplitude is built. The interpretation of ${\rm Im}\,\Gamma_{\text{eff}}$ as a vacuum-decay probability therefore does not carry over to an expanding spacetime. The appropriate framework is instead the Schwinger--Keldysh formalism introduced in \cref{sec:Pathintegral}, in which physical observables are evaluated as expectation values in a given initial state,
\begin{equation}
    \langle {\hat O}(x)\rangle_{\rm in}
    =
    \langle 0_{\rm in}|U^\dagger(\tau,\tau_i)\,{\hat O}_I(x)\,U(\tau,\tau_i)|0_{\rm in}\rangle .
\end{equation}
Here ${\hat O}_I(x)$ denotes the operator corresponding to the physical observable ${\hat O}$ in the interaction picture, while $U(\tau,\tau_i)$ is the interaction-picture time-evolution operator from the initial time $\tau_i$ to the observation time $\tau_f$. The in--in formulation thus allows observables such as the induced current and stress tensor to be computed directly in the  $\ket{0_\text{in}}$ state, without reference to an out-vacuum.

\subsection{Fermionic CTP influence functional}
\label{sec:influence_functional}

Integrating out the charged fermions yields a non-local closed-time-path effective action for the gauge field, and more generally for any coupled background fields, including the spacetime geometry. Although the fermions are initially taken to be in their vacuum rather than in a thermal or populated state, their vacuum fluctuations can still induce quantum correlations, dissipation, noise, and decoherence in the resulting effective dynamics. The influence of the fermionic sector is encoded in the Feynman--Vernon influence functional \cite{Calzetta:1993qe}, $S_{\text{IF}}$. In this language, the CTP partition function can be written as 
\begin{align}
\label{eq:Z_QED_split}
    \mathcal{Z}&=
    \int \mathcal{D}A_\mu^\pm\ \rho_A[A_\mu^+,A_\mu^-]\, \exp\left\{i\left[ S_{\rm A}[A_\mu^+]-S_{\rm A}[A_\mu^-]+\int d^4x\,(J^{\mu +}A_\mu^+-J^{\mu -}A_\mu^-)\right]\right\}\nonumber\\
    &\quad\times \exp\left\{iS_{\rm IF}[A_\mu^+,A_\mu^-]\right\},
\end{align}
where the action,
\begin{equation}
    S_{\rm A}[A_\mu]=\int d^4x\left[-\frac{1}{4}F_{\mu\nu}F^{\mu\nu}-\frac{1}{2\xi}(\partial_\mu A^{\mu})^2\right],
\end{equation}
describes the gauge-fixed Maxwell sector appearing in \cref{eq:QED_gf} and the fermionic influence functional is defined by
\begin{align}
\label{eq:influence_functional_def}
  e^{iS_{\rm IF}}
   & =
    \int_{|0_{\text{in}}\rangle} \mathcal{D}\bar{\psi}^\pm \mathcal{D}\psi^\pm\,
    \exp\left\{
        i\left[\int d^4x\ 
        \bar{\psi}^+
        (i\slashed{\partial}-\mathfrak{q}\slashed{A}^+)
        \psi^+     -
        \bar{\psi}^-
        (i\slashed{\partial}-\mathfrak{q}\slashed{A}^-)
        \psi^-\right]
    \right\}.
\end{align}

In the semi-classical approximation, the CTP effective action for the Maxwell degrees of freedom is then
\begin{equation}
\label{eq:Gamma_QED_oneloop}
    \Gamma_{\text{eff}}[A_\mu^+,A_\mu^-]\approx S_{\rm A}[A_\mu^+]-S_{\rm A}[A_\mu^-]+S_{\text{IF}}[A_\mu^+,A_\mu^-].
\end{equation}
At quadratic order, the effective action captures the corrections to the photon dynamics and associated decoherence effects, while higher-order terms generate non-linear interactions induced by the fermion loops. Since the Maxwell action is real, the imaginary part of $\Gamma_{\text{eff}}$ is entirely determined by the influence functional.

The physical interpretation of $S_{\rm IF}$ is particularly transparent in the operator formulation. For an initial fermionic data given by the density matrix $\hat{\rho}_\psi(\tau_i)$, one may write
\begin{equation}
\label{eq:trace}
    e^{iS_{\rm IF}[A_\mu^+,A_\mu^-]}
    =\Tr_{\psi}\left[\hat{U}[A_\mu^+]\,\hat{\rho}_\psi(\tau_i)\,\hat{U}^{\dagger}[A_\mu^-]\right],
\end{equation}
where
\begin{equation}
    \hat{U}[A_\mu]\equiv\hat{U}_A(\tau_f,\tau_i)=\mathcal{T}\exp\left[i\int d^4x\,A_\mu(x)\hat{j}^{\mu}(x)\right]
\end{equation}
is the time evolution operator and $A_\mu\hat j^\mu$ is the interacting Hamiltonian density. For the pure initial state, $\hat{\rho}_\psi(\tau_i\to-\infty)=|0_{\text{in}}\rangle\langle 0_{\text{in}}|$, \cref{eq:trace} reduces to
\begin{equation}
    \label{eq:IF_operator_form_pure}
    \exp\left\{\,iS_{\rm IF}[A_\mu^+,A_\mu^-]\right\} = \langle 0_{\text{in}}|U^\dagger[A_\mu^-]\,U[A_\mu^+]|0_{\text{in}}\rangle .
\end{equation}
The influence functional therefore measures the overlap between the fermionic states generated by the two gauge-field histories. When these histories become distinguishable through their imprint on the fermionic sector, the magnitude of this overlap is reduced, giving rise to an imaginary contribution to the effective action.

For the stochastic backgrounds considered here, we further average over the statistical state of the gauge field,
\begin{equation}
\label{eq:Gamma_eff_def}
    e^{\,iS_{\text{IF}}[A_\mu^+,A_\mu^-]}
    =\left\langle\hspace{-0.2cm}\left\langle\widetilde{\mathcal{T}}\exp\!\left(-i\!\int d^4x\;A^-_\mu\hat{j}^{\mu}\right)\mathcal{T}\exp\!\left(i\!\int d^4x\;A^+_\mu\hat{j}^{\mu}\right)\!\right\rangle\hspace{-0.2cm}\right\rangle
\end{equation}
with
\begin{equation}
    \llangle \cdots \rrangle
    \equiv \big\langle \langle \cdots \rangle_{\rm BD} \big\rangle_{\rm st},
    \quad \text{and} \quad
    \langle \cdots \rangle_{\rm BD}\equiv \langle 0_{\rm in}|\cdots|0_{\rm in}\rangle,
\end{equation}
where $\langle\cdots\rangle_{\rm BD}$ denotes the quantum expectation value in the Bunch--Davies vacuum of the fermionic sector. We now expand $S_{\mathrm{IF}}$ perturbatively in the coupling $\mathfrak{q}$. Owing to the stochastic nature of the gauge-field background, the linear contribution vanishes upon statistical averaging, and the leading contribution is therefore quadratic in the gauge field. To second order in $A_\mu$, we then obtain
\begin{equation}
\label{eq:Gamma2_CTP}
    S_{\text{IF}}
    =\frac{i}{2}\int d^4x\int d^4y\;\left\langle \begin{pmatrix}    A_\mu^+(x) & \hspace{0.2cm}  -A_\mu^-(x) \end{pmatrix}\,\boldsymbol{G}^{\mu\nu}(x,y)\, \begin{pmatrix}A_\nu^+(y) \\ -A_\nu^-(y) \end{pmatrix}\right\rangle_{\text{st}}+\mathcal{O}(\mathfrak{q}^3).
\end{equation}
where $x\equiv(\tau,\bx)$ and $y\equiv(\tau',\by)$ and the matrix $\boldsymbol{G}^{\mu\nu}$ contains the four current-current correlators,
\begin{equation}
    \boldsymbol{G}^{\mu\nu}(x,y)
    \equiv
    \begin{pmatrix}
    G_{++}^{\mu\nu}(x,y) & G_{+-}^{\mu\nu}(x,y)\\[4pt]
    G_{-+}^{\mu\nu}(x,y) & G_{--}^{\mu\nu}(x,y)
    \end{pmatrix}
    =
    \begin{pmatrix}
    \langle \mathcal{T}\,\hat{j}^\mu(x)\hat{j}^\nu(y)\rangle_{\text{BD}} & \langle\hat{j}^\nu(y)\hat{j}^\mu(x)\rangle_{\text{BD}}\\[4pt]
    \langle\hat{j}^\mu(x)\hat{j}^\nu(y)\rangle_{\text{BD}} & \langle\widetilde{\mathcal{T}}\,\hat{j}^\mu(x)\hat{j}^\nu(y)\rangle_{\text{BD}}
    \end{pmatrix}.
    \label{eq:G-}
\end{equation}
Notably, the time- and anti-time ordered correlators are related to the Wightman functions through
\begin{equation}
\label{eq:Scwhinger_rel}
    G_{++}^{\mu\nu}(x,y)+G_{--}^{\mu\nu}(x,y)=G_{+-}^{\mu\nu}(x,y)+G_{-+}^{\mu\nu}(x,y).
\end{equation}
\begin{figure}[ht]
    \centering
    \includegraphics[width=0.8\linewidth,trim=12cm 12cm 12cm 12cm,clip]{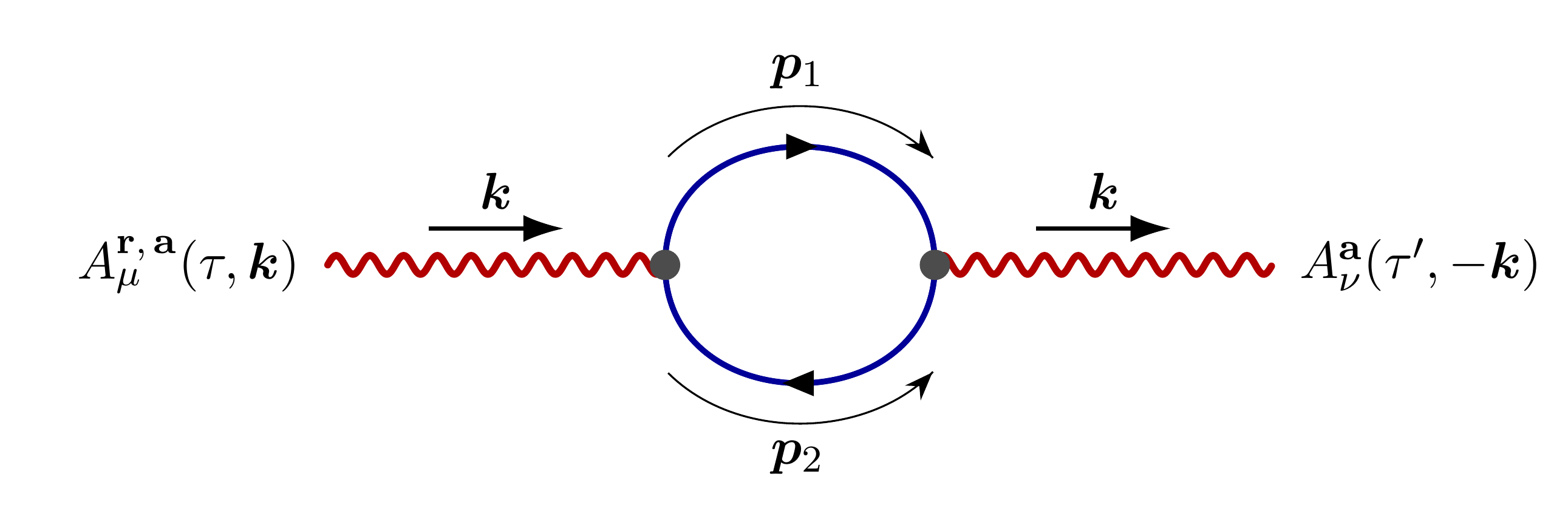}
    \caption{Leading one-loop contribution to the influence functional.
Its real and imaginary parts give rise to the causal and
noise kernels in \cref{eq:Gamma2_Keldysh}, respectively.}
    \label{fig:Feynman}
\end{figure}

It is now useful to pass to the Keldysh basis, where we define
\begin{equation}
\label{eq:Keldysh_basis}
    A^{\mathbf{r}}_\mu = \tfrac{1}{2}(A^+_\mu+A^-_\mu),
    \quad\text{and}\quad
    A^\mathbf{a}_\mu = A^+_\mu-A^-_\mu,
\end{equation}
with late-time endpoints given by
\begin{equation}
    A_\mu^{\mathbf{r}}(\tau_f,\bx)=A_\mu(x),
    \quad\text{and}\quad
    A_\mu^{\mathbf{a}}(\tau_f,\bx)=0.
\end{equation}
Now, using \cref{eq:Scwhinger_rel} the current-current correlators reorganise into
\begin{align}
   % \boldsymbol{G}^{\mu\nu}(x,y)   &\mapsto
    \begin{pmatrix}
    G_{\mathbf{r}\mathbf{r}}^{\mu\nu}(x,y) & G_{\mathbf{r}\mathbf{a}}^{\mu\nu}(x,y)\\[4pt]
    G_{\mathbf{a}\mathbf{r}}^{\mu\nu}(x,y) & G_{\mathbf{a}\mathbf{a}}^{\mu\nu}(x,y)
    \end{pmatrix} %\nonumber\\ &
    =
    \begin{pmatrix}
     0 & \Theta(\tau'-\tau)\langle[\hat{j}^\mu(x),\hat{j}^\nu(y)]\rangle_{\text{BD}}\\[4pt]
    -\Theta(\tau-\tau')\langle[\hat{j}^\mu(x),\hat{j}^\nu(y)]\rangle_{\text{BD}} & \langle\{\hat{j}^\mu(x),\hat{j}^\nu(y)\}\rangle_{\text{BD}}
    \end{pmatrix},
\end{align}
where $\Theta(x)$ is the Heaviside step function. The influence action then separates into causal and noise terms,
\begin{equation}
\label{eq:Gamma2_Keldysh}
    S_{\text{IF}}
    =
    \int d^4x\!\int d^4y\ \left[
    \langle A^{\mathbf{a}}_\mu(x)\,\Pi^{\mu\nu}(x,y)\,A^{\mathbf{r}}_\nu(y)\rangle_{\text{st}}
    +
    \frac{i}{2}\langle A^{\mathbf{a}}_\mu(x)\,N^{\mu\nu}(x,y)\,A^{\mathbf{a}}_\nu(y)\rangle_{\text{st}}\right]+\mathcal{O}(\mathfrak{q}^3),
\end{equation}
with the respective kernels
\begin{equation}
    \Pi^{\mu\nu}(x,y)=i\Theta(\tau-\tau')\langle[\hat{j}^\mu(x),\hat{j}^\nu(y)]\rangle_{\text{BD}} \quad \text{and}\quad N^{\mu\nu}(x,y)=\frac{1}{2}\langle\{\hat{j}^\mu(x),\hat{j}^\nu(y)\}\rangle_{\text{BD}}.
    \label{eq:Pi-N}
\end{equation}
The first term is real and causal, and describes the causal fermionic response to the gauge field. The second term is purely imaginary and encodes fluctuations, decoherence, and particle-production effects. Diagrammatically, these contributions correspond to the real and imaginary components of the one-loop diagram shown in  \cref{fig:Feynman}.

To evaluate the leading contribution to the effective action, we can express the causal and noise kernels in terms of free fermionic two-point functions. Since the calculation is perturbative, we can use Wick’s theorem to write the current-current correlator appearing in $S_{\text{IF}}$ as products of Bunch--Davies Wightman functions. Defining
\begin{equation}
\label{eq:Wightman}
    \boldsymbol{S}^{>}(x,y) = \langle\hat{\psi}(x)\hat{\bar{\psi}}(y)\rangle_{\text{BD}},
    \quad\text{and}\quad
    \boldsymbol{S}^{<}(x,y) = -\langle\hat{\bar{\psi}}(y)\hat{\psi}(x)\rangle_{\text{BD}},
\end{equation}
we obtain
\begin{equation}
    \Pi^{\mu\nu}(x,y)
    =-i\mathfrak{q}^2\Theta(\tau-\tau')\operatorname{tr}\!\left[\gamma^\mu \boldsymbol{S}^{>}(x,y)\,\gamma^\nu \boldsymbol{S}^{<}(y,x)-\gamma^\nu \boldsymbol{S}^{>}(y,x)\,\gamma^\mu \boldsymbol{S}^{<}(x,y)\right]
\end{equation}
and
\begin{equation}
\label{eq:noise_Wightman}
    N^{\mu\nu}(x,y)=-\frac{\mathfrak{q}^2}{2}\operatorname{tr}\!\left[\gamma^\mu \boldsymbol{S}^{>}(x,y)\,\gamma^\nu \boldsymbol{S}^{<}(y,x) + \gamma^\nu \boldsymbol{S}^{>}(y,x)\,\gamma^\mu \boldsymbol{S}^{<}(x,y)\right],
\end{equation}
where the traces are over spinor indices. Therefore, the imaginary part of the effective action can be evaluated directly from the trace in \cref{eq:noise_Wightman}. We now turn to its computation.

\subsection{Quantisation of the massless Dirac field}\label{sec:Q-massless}

We now quantise the free massless Dirac field on the $dS_4$ background. Here, we focus on the canonically normalised field defined in \cref{eq:CN-field}. The field operator admits the mode expansion
\begin{equation}
    \hat{\psi}(\tau,\bx)=\int d^3\bk\ \hat{\psi}_{\bk}(\tau)\, e^{i\bk\cdot\bx},
\end{equation}
where
\begin{equation}
    \hat{\psi}_{\bk}(\tau)=\sum_{s=\pm}\left[\hat{a}_{s,\bk}U_{s,\bk}(\tau)+\hat{b}^{\dagger}_{s,-\bk}V_{s,-\bk}(\tau)\right]
\end{equation}
and the sum running over the helicity states $s=\pm$. The creation and annihilation operators satisfy
\begin{equation}
\label{eq:Anticommab-}
    \{\hat{a}_{s,\bk},\hat{a}^{\dagger}_{s',\bk'}\}=\{\hat{b}_{s,\bk},\hat{b}^{\dagger}_{s',\bk'}\}=\delta_{ss'}\delta^{(3)}(\bk-\bk'),
\end{equation}
with all remaining anticommutators vanishing. This in turn implies a Fock basis construction defined by
\begin{equation}
    \hat a_{s,\bk}|0_{\text{in}}\rangle=\hat b_{s,\bk}|0_{\text{in}}\rangle=0.
\end{equation}

The particle and antiparticle mode functions can be written in a helicity basis as
\begin{equation}
\label{eq:UVpsipls}
    U_{s,\bk}(\tau)
    =
    \frac{1}{\sqrt{2}}
    \begin{pmatrix}
        {\bf E}_s u^{\uparrow}_s(\tau,k)\\
    s{\bf E}_s u^{\downarrow}_s(\tau,k)
    \end{pmatrix},
    \qquad
    V_{s,-\bk}(\tau)
    =
    \frac{1}{\sqrt{2}}
    \begin{pmatrix}
        {\bf E}_s v^{\uparrow}_s(\tau,k)\\
        s{\bf E}_s v^{\downarrow}_s(\tau,k)
    \end{pmatrix},
\end{equation}
where ${\bf E}_s$ are helicity eigenspinors satisfying ${\bf E}_s^\dagger{\bf E}_{s'}=\delta_{ss'}$. Using charge-conjugation symmetry, we find
\begin{equation}
\label{eq:uvpl}
    u^{\uparrow}_{s}(\tau,k)=\bar{v}^{\uparrow}_{s}(\tau,k),\qquad u^{\downarrow}_{s}(\tau,k)=-\bar{v}^{\downarrow}_{s}(\tau,k).
\end{equation}
The antiparticle modes are therefore fixed by the particle modes, so it is sufficient to solve
\begin{equation}
    \begin{cases}
        i\partial_{\tau}u^{\uparrow}_s-k u^{\downarrow}_s&=0,\\[4pt]
        i\partial_{\tau}u^{\downarrow}_s-k u^{\uparrow}_s&=0.
    \end{cases}
\end{equation}
We then arrive at
\begin{equation}
    (\partial_\tau^2+k^2)\,u^{\uparrow,\downarrow}_s(\tau,k)=0.
\end{equation}
The Bunch--Davies condition selects the positive-frequency solution
\begin{equation}
    \label{eq:BD}
    u^\uparrow_s(\tau,k)
    =
    u^\downarrow_s(\tau,k)
    = \frac{1}{(2\pi)^{\frac32}}
    e^{-ik\tau}.
\end{equation}
With the decomposition in \cref{eq:UVpsipls}, the canonical normalisation condition is
\begin{equation}
\frac12
\left(
|u^\uparrow_s|^2
+
|u^\downarrow_s|^2
\right)
=
\frac{1}{(2\pi)^3},
\end{equation}
which is satisfied by \cref{eq:BD}.

To evaluate the noise kernel in \cref{eq:noise_Wightman}, we require the fermionic greater and lesser Wightman functions. The physical propagators and the propagators of the rescaled field are related by
\begin{equation}
\label{eq:physical_rescaled_propagator}
    S^{\gtrless}_{\mathrm{phys}}(x,y)
    =
    \frac{1}{[a(\tau)a(\tau')]^{\frac32}}\,
    S^{\gtrless}(x,y),
\end{equation}
where $S(x,y)$ denotes the rescaled propagator.  We also use the Fourier conventions
\begin{align}
    S^{\gtrless}(x,y)=\int d^3\bk\ S^{\gtrless}(\tau,\tau',\bk)\, e^{\pm i\bk\cdot(\by-\bx)}.
    \label{eq:S-gtrless}
\end{align}
Under these, the rescaled Wightman functions mimic the flat-space definitions used in standard QFT references \cite{Peskin:1995ev,Schwartz:2014sze}. The greater Wightman function is defined as
\begin{align}
S^>(\tau,\tau';\bk)
=
\sum_{s=\pm}
U_{s,\bk}(\tau)\bar U_{s,\bk}(\tau'),
\label{eq:greaterWsh}
\end{align}
The lesser Wightman function is defined by $V_{s,\bk}(\tau)$ as $S^<(\tau,\tau';\bk)
=-
\sum_{s=\pm}
V_{s,\bk}(\tau)\bar V_{s,\bk}(\tau')$. Further details are provided in \cref{app:Wightman}.
The matrix structure of the propagator is constrained by the symmetries of the background which after substituting the Bunch--Davies mode functions \cref{eq:BD} gives
\begin{equation}
\label{eq:GreaterW_modes}
S^{\gtrless}(\tau,\tau';\bk)
= \mp \frac{1}{(2\pi)^3}
\frac{\gamma^\mu{K}_\mu}{2k}
e^{\mp ik(\tau-\tau')},
\end{equation}
where ${K}_\mu=(-k,\bk)$ is a null four-vector. The vector $K^\mu$ is introduced only as a compact way of writing the spinor projector. It is not the external momentum appearing in the loop kernel below.

\subsection{Noise kernel and the stochastic Schwinger effect}\label{sec:Noise-Stoch}

Having derived the  Wightman functions, we now compute the correlator entering \cref{eq:noise_Wightman}. The resulting kernel determines the imaginary part of the effective action through
\begin{equation}
\label{eq:IM_action}
    \mathrm{Im}\,\Gamma_{\rm eff}^{}[A^{\mathbf{a}}_\mu]
    = \frac12
    \int d\tau\,d\tau'
    \int d^3\bk\;
    N_{\bk}^{\mu\nu}(\tau,\tau')
    \langle
    A^{\mathbf{a}}_\mu(\tau,\bk)\,
    A^{\mathbf{a}}_\nu(\tau',-\bk)
    \rangle_{st} +\mathcal{O}(\mathfrak{q}^3),
\end{equation}
where
\begin{equation}
\label{eq:noise_kernel_definition}
    N_{\bk}^{\mu\nu}(\tau,\tau')
    \equiv
    -\frac{\mathfrak{q}^2}{2}\int d^3\bp_1\;
    \mathrm{tr}
    \!\left[
    \gamma^\mu
    S^>(\tau,\tau';\bp_1)\,
    \gamma^\nu
    S^<(\tau',\tau;\bk-\bp_1)
    \right]+\text{h.c.}
\end{equation}
with $``\text{h.c.}"$ denoting Hermitian conjugation and the reality condition $A_\mu^*(\tau,\bk)=A_\mu(\tau,-\bk)$. Momentum conservation in the loop fixes the second internal momentum to be $\bp_2=\bk-\bp_1$ (see Appendix~\cref{App:momentum_structure} and  \cref{fig:Feynman}), while gauge invariance implies transversality of the kernel 
\begin{equation}
\label{eq:Ward}
    \partial_\mu^{(x)}N^{\mu\nu}(x,y)=0,
    \quad
    \text{and}
    \quad
    \partial_\nu^{(y)}N^{\mu\nu}(x,y) =0.
\end{equation}
Note that $N^{\mu\nu}(x,y)$ is the conformally rescaled kernel written with flat conformal indices.\footnote{For the physical curved-space current kernel, $N_{\text{phys}}^{\mu\nu}(x,y)$, the Ward identity instead takes the covariant form $\nabla_\mu^{(x)}N_{\text{phys}}^{\mu\nu}(x,y)=0$,
and $\nabla_\nu^{(y)}N_{\text{phys}}^{\mu\nu}(x,y)=0
$.}

Substituting the fermionic Wightman functions and using the trace identities \cref{eq:tr-1} and \cref{eq:tr-2} gives
\begin{equation}
\label{eq:Nk-munu}
    N_{\bk}^{\mu\nu}(\tau,\tau')
    =\mathfrak{q}^2\int
    \frac{d^3\bp_1}{p_1p_2}\,
    \Big[
    P_1^\mu P_2^\nu
    +
    P_1^\nu P_2^\mu
    -
    (P_1\!\cdot\!P_2)\eta^{\mu\nu}
    \Big]
    \cos\!\big[(p_1+p_2)(\tau-\tau')\big]
    \Bigg|_{\bp_2=\bk-\bp_1},
\end{equation}
where $P_{A}^\mu=(p_{A},\bp_{A})$ with $A\in\{1,2\}$ are null vectors for the internal momenta. The remaining momentum integral is evaluated in Appendix~\cref{subsec:momentum-integrals} and gives
\begin{equation}
\label{eq:noise_spectral}
    N_{\bk}^{\mu\nu}(\tau,\tau')= \frac{\pi \mathfrak{q}^2}{3}\int^{\infty}_{ k}d\omega\, \left(K^\mu K^\nu-K^2\eta^{\mu\nu}\right)\,[e^{-i\omega(\tau-\tau')}+e^{i\omega(\tau-\tau')}],
\end{equation}
with $K^\mu=(\omega,\bk)$. Here, we have used the spectral decomposition of the cosine
\begin{equation}
\label{eq:cos_spectral}
    \cos[(p_1+p_2)(\tau-\tau')]
    =
    \frac12
    \int_{-\infty}^{\infty}
    d\omega\;
    e^{-i\omega(\tau-\tau')}
    \Big[
    \delta(p_1+p_2-\omega)
    +
    \delta(p_1+p_2+\omega)
    \Big].
\end{equation}
From the above, we see that the kernel is supported only for timelike momenta, $|\omega|\ge k$, and additionally is transverse,
\begin{equation}
    K_\mu\left(K^\mu K^\nu-K^2\eta^{\mu\nu}\right)=0,
\end{equation}
as demanded by the Ward identities \cref{eq:Ward}. The timelike support is a remnant of the flat-space kinematic threshold for producing a fermion-antifermion pair, which is made manifest through the conformal invariance of the theory.

Although $\omega$ is the variable conjugate to the conformal-time separation, it should not in general be interpreted as a conserved physical frequency of the stochastic gauge background, since de Sitter spacetime does not possess the corresponding global time-translation symmetry. We therefore return to the time domain and let powers of $\omega$ act as time derivatives on the gauge fields. The transverse tensor structure in \cref{eq:noise_spectral} then reorganises into the field strength contraction
\begin{equation}
\label{eq:kernel_FF}
    N_{\bk}^{\mu\nu}(\tau,\tau')
    \langle
    A^{\mathbf{a}}_\mu(\tau,\bk)
    A^{\mathbf{a}}_\nu(\tau',-\bk)
    \rangle_{st}
    = 
    \frac{\mathfrak{q}^2\pi}{3}\,
    \mathcal I_{k}(\tau-\tau')\,
    \langle
    -\tfrac{1}{2}F^{\mathbf{a}}_{\mu\nu}(\tau,\bk)
    F^{\mu\nu\, \mathbf{a}}(\tau',-\bk)
    \rangle_{\text{st}},
\end{equation}
where
\begin{equation}
\label{eq:Ikernel}
    \mathcal I_k(\tau-\tau')\equiv 2\left\{\pi\delta(\tau-\tau')-\frac{\sin[k(\tau-\tau')]}{\tau-\tau'}\right\}.
\end{equation}
As expected, $\text{Im}\Gamma_{\text{eff}}$ depends exclusively on a gauge invariant two-point stochastic average. Substituting \cref{eq:kernel_FF} into \cref{eq:IM_action} gives
\begin{equation}
\label{eq:ImGamma_final}
    \mathrm{Im}\Gamma_{\text{eff}}
    = 
    \frac{\mathfrak{q}^2\pi}{6}
    \int d\tau d\tau'\!\int d^3\bk\ 
  \mathcal{I}_k(\tau-\tau') \, 
    \langle 
    -\tfrac{1}{2}F^{\mathbf{a}}_{\mu\nu}(\tau,\bk)
    F^{\mu\nu\,{\mathbf{a}}}(\tau',-\bk)
    \rangle_{\text{st}} +\mathcal{O}(\mathfrak{q}^3).
\end{equation}
Most notably, no explicit power of $H$ appears in \cref{eq:ImGamma_final}. This is the loop-level signature of the conformal argument in \cref{sec:conf}. The absence of explicit $H$-dependence in the leading noise kernel reflects the conformal mapping of the massless fermionic dynamics to Minkowski space. Quantum conformal invariance is broken by the trace anomaly and the running of the gauge coupling. The anomalous Weyl dependence, however, contributes a local real term  and therefore does not modify the imaginary part considered here. The two terms in the scalar kernel \cref{eq:Ikernel} play distinct roles. The first is local in conformal time, whereas the sinc term provides a non-local memory contribution, coupling the gauge field at different times. Together, the two terms encode the temporal non-locality of the fermionic response and select the timelike modes capable of producing fermion pairs.

The field-strength contraction separates the electric and magnetic
contributions as
\begin{equation}
    \langle -\tfrac{1}{2}F_{\mu\nu}^{{\mathbf{a}}}(\tau,\bk)
    F^{\mu\nu\,{\mathbf{a}}}(\tau',-\bk)\rangle_{\text{st}}
    =
    \left\langle
    E_i^{\mathbf{a}}(\tau,\bk)E_i^{\mathbf{a}}(\tau',-\bk)
    -
    B_i^{\mathbf{a}}(\tau,\bk)B_i^{\mathbf{a}}(\tau',-\bk)
    \right\rangle_{\text{st}},
\end{equation}
with $E_i^{\mathbf{a}}=F_{0i}^{\mathbf{a}}$ and $B_i^{\mathbf{a}}=\frac12\epsilon_{ijk}F_{jk}^{\mathbf{a}}$. At this perturbative order the kernel has support only on timelike Fourier components of the gauge background. These correspond to electric-like field configurations, for which $-F_{\mu\nu}F^{\mu\nu}>0$. This logic mimics the usual flat-space intuition, where only electric fields can do work on charged particles and magnetic-like configurations do not lead to Schwinger pair production. We also see that in the flat space limit $a(\tau)\to1$ the result in \cref{eq:ImGamma_final} reproduces the flat-space result \cite{VicenteGarcia-Consuegra:2025lkh}, see \cref{sec:frequency_resolved_limit} for the explicit calculation.

\subsection{Extension to FLRW backgrounds}
\label{sec:generic_FLRW}
The results of this section are not specific to de Sitter spacetime. The essential ingredients in \hyperref[sec:influence_functional]{\namecrefs{sec:influence_functional}~\ref*{sec:influence_functional}} through \ref{sec:Noise-Stoch} are the conformal flatness of the background and the classical conformal invariance of the massless fermionic sector, both of which hold for any spatially flat FLRW geometry
\begin{equation}
    ds^2=a^2(\tau)(-d\tau^2+d\bx^2),
\end{equation}
with $a(\tau)$ an arbitrary scale factor.\footnote{This is the physically relevant case, since inflation drives the spatial curvature towards zero and the observed Universe is spatially flat to within a fraction of a percent \cite{Planck:2018jri}.} The plane-wave harmonics of such a geometry are precisely those used in the mode decomposition of \cref{sec:Q-massless}.

With this choice, the field redefinitions in \cref{eq:CN-field} remove the scale factor from the massless Dirac action, and the vector current density reduces to its Minkowski form exactly as in \cref{sec:conf}. The physical current is then
\begin{equation}
    j^\mu_{\rm phys}(x)=a^{-4}(\tau)\,j^\mu(x),
\end{equation}
while $\sqrt{-g}=a^4(\tau)$. The combination entering the influence functional therefore carries no scale factors,
\begin{equation}
    \sqrt{-g(\tau)}\,\sqrt{-g(\tau')}\,
    j^\mu_{\rm phys}(\tau,\bx)\,
    j^\nu_{\rm phys}(\tau',\by)
    =
    j^\mu(\tau,\bx)\, j^\nu(\tau',\by),
\end{equation}
and identical to its Minkowski counterpart. 

The kernel $\mathcal I_k(\tau-\tau')$ in \cref{eq:Ikernel} is generated entirely by the rescaled mode functions (see \cref{eq:Ikernelapp}), which coincide with their flat-space counterparts, so the loop integral producing it is insensitive to the expansion history. The influence functional, the rescaled current correlator and the noise kernel therefore carry over unchanged, and \cref{eq:ImGamma_final} applies to any spatially flat FLRW background, whether during inflation, reheating or radiation domination, with the cosmological model entering only through the stochastic input $\langle F_{\mu\nu}F^{\mu\nu}\rangle_{\rm st}$.

\section{Consistency and limits of the stochastic description}\label{sec:phenomenological_production}

Before turning to further applications, it is useful to clarify the consistency conditions and limiting regimes of the stochastic description developed above. We first show that the massless fermionic result remains infrared safe in de Sitter, despite the potential sensitivity of massless fields to long-wavelength modes. We then discuss the conditions under which the prescribed stochastic gauge background can consistently be treated as classical while remaining a spectator of the cosmological geometry. Finally, we examine the long-wavelength limit of the stochastic result and its relation to the conventional static Schwinger effect, emphasising the distinction between the perturbative stochastic description and the non-perturbative constant-field problem.

\subsection{Infrared safety of massless QED in de Sitter}\label{sec:IR}

A natural question in the massless limit is whether de Sitter expansion induces an infrared enhancement of the fermionic sector that could render the production kernel singular. This concern is particularly relevant in de Sitter, where massless minimally coupled scalars exhibit the well-known accumulation of superhorizon fluctuations \cite{PhysRevD.31.710,Allen:1987tz}. Massless fermions behave differently. Like conformally coupled massless scalars, they are conformally invariant in four dimensions and therefore do not undergo superhorizon freezing or develop an intrinsic de Sitter infrared enhancement \cite{Birrell:1982ix}. The massless limit consequently removes the finite threshold for pair production without introducing an intrinsic de Sitter infrared enhancement in the fermionic correlators. This property can also be understood directly from the effective action obtained after integrating out the massless fermions \cite{Peskin:1995ev,Itzykson:1980rh,Schwartz:2014sze,Donoghue:2015xla}. After the Weyl rescaling, the non-local fermionic kernel is exactly the corresponding flat-space kernel expressed in conformal coordinates, up to local counterterms and anomaly-induced local terms. In particular, no additional infrared scale proportional to $H$ is generated by the fermionic correlator.

The absorptive part of the quadratic effective action is given by \cref{eq:ImGamma_final}. In the infrared, its momentum-space integrand behaves schematically as
\begin{equation}
\label{eq:limFF}
  \lim_{k\rightarrow 0}   \, k^3 
  \, 
    \langle 
    -\tfrac{1}{2}F^{\mathbf{a}}_{\mu\nu}(\tau,\bk)
    F^{\mu\nu\,{\mathbf{a}}}(\tau,-\bk)
    \rangle_{\text{st}}.
\end{equation}
where the factor $k^3$ arises from the three-dimensional momentum measure. Hence, provided the stochastic gauge-field correlator itself is infrared regular, the small-$k$ contribution remains finite. Any infrared singularity must therefore originate from the prescribed gauge-field spectrum rather than from the massless fermion loop. In particular, the absence of terms scaling as $H/k$ or $H^2/k^2$ shows that de Sitter expansion does not induce an intrinsic infrared enhancement in the massless fermionic sector. Although the pair-production threshold becomes gapless for $m=0$, the absorptive kernel remains regular in the long-wavelength limit.

\subsection{Classicality of the stochastic gauge background}\label{eq:class-BG}

There is also an important consistency requirement on the amplitude of the gauge background. Throughout our analysis, the gauge field is treated as a prescribed classical stochastic configuration rather than as a vacuum fluctuation. We assume that its spectrum is peaked on a sub-horizon scale,
\begin{equation}
    k_{\rm peak}\gtrsim a(\tau)H.
\end{equation}
Classicality then requires a large occupation number around the peak,
\begin{equation}
    n_A(k_{\rm peak})\gg1,
\label{eq:classicality_occupation}
\end{equation}
where $n_A$ denotes the occupation number of the gauge-field. The field amplitude must lie parametrically above the vacuum fluctuations at the characteristic scale of the background. So that the stochastic gauge field entering \cref{eq:ImGamma_final} is parametrically larger than the inflationary vacuum fluctuations and may consistently be regarded as a classical background.  At the same time, neglecting the backreaction of the gauge sector on the inflationary geometry requires that the energy density of the gauge field be subleading comparing to the total energy scale of inflation.

For a statistically homogeneous and isotropic stochastic gauge background, the ensemble-averaged physical energy density is spatially uniform,
\begin{equation}
    \left\langle \varrho_A(\tau,\bx)\right\rangle_{\rm st} =
    \frac12
    \left\langle
    E^2+B^2
    \right\rangle_{\rm st}.
\end{equation}
Requiring the gauge field to lie above the characteristic inflationary vacuum-fluctuation scale while remaining a spectator of the inflationary background then defines the broad hierarchy
\begin{equation}
    H^4
    \ll
    \varrho_A
    \ll
    M_{\rm Pl}^2H^2.
\label{eq:classical_window}
\end{equation}
The lower bound characterises a classical stochastic background, while the upper bound ensures negligible backreaction on the inflationary geometry.

For later comparison with the Fourier-space expression entering \cref{eq:ImGamma_final}, it is useful to translate this hierarchy  to the field-strength correlator. For a spectrum concentrated around $k_{\rm peak}$, and assuming an electric-like configuration without a  large cancellation between the electric and magnetic contributions, one has schematically
\begin{equation}
    \varrho_A
    \sim
    \left.k^3\left|\left\langle -\tfrac12 F_{\mu\nu}(\tau,\bk)F^{\mu\nu}(\tau,-\bk)\right\rangle_{\rm st}^{\prime}\right|\right|_{k\simeq k_{\rm peak}},
\end{equation}
where the \textit{prime} denotes that the momentum-conserving delta function has been stripped off. This relation should be understood parametrically, since the energy density involves $(E^2+B^2)$, whereas the invariant entering the fermionic kernel probes $(E^2-B^2)$. Roughly speaking, the hierarchy in \cref{eq:classical_window} therefore translates, into
\begin{equation}
    H^4
    \ll
    \left.k^3\left|\left\langle -\tfrac12F_{\mu\nu}(\tau,\bk)F^{\mu\nu}(\tau,-\bk)\right\rangle_{\rm st}^{\bold{\prime}}\right|\right|_{k\simeq k_{\rm peak}}
    \ll
    M_{\rm Pl}^2H^2,
\label{eq:classical_window_F}
\end{equation}
for time separations $|\tau-\tau'|$ within the coherence time of the stochastic background.

\subsection{Long-wavelength limit and connection to static Schwinger effect}\label{sec:constant_field_limit}
It is instructive to compare the long-wavelength limit of the stochastic result \cref{eq:ImGamma_final} with the strictly static Schwinger effect  \cref{eq:dS-Upsilon-m0}. For a spatially homogeneous electric configuration, with $B=0$ and a time-independent field amplitude, the Fourier support is localised at $\bk=0$,
\begin{equation}
    -\frac12 F_{\mu\nu}(\tau,\bk)F^{\mu\nu}(\tau',-\bk)
    \longrightarrow
    \frac{V_3 \, E^2}{(2\pi)^3}\,\delta^{(3)}(\bk),
\end{equation}
where $V_3\equiv\int d^3\bx=(2\pi)^3\delta^{(3)}(\boldsymbol 0)$. Substituting the homogeneous limit into the stochastic decay rate obtained in \cref{eq:ImGamma_final} gives
\begin{equation}
    \left.2{\rm Im}\Gamma_{\rm eff}^{(2)}\right|_{\text{stochastic};\, \bk\rightarrow0}
    =
    \frac{\mathfrak q^2E^2 \, V_4}{12\pi} \quad\text{(perturbative)}.
\label{eq:stochastic_soft_limit}
\end{equation}
By contrast, a strictly constant electric background is described by the full Schwinger result,
\begin{equation}
    \left.2{\rm Im}\Gamma_{\rm eff}^{\textcolor{white}{(2)}}
    \right|_{\rm static}
    =
    \frac{\mathfrak q^2E^2 \, V_4}{4\pi^3}
    \sum_{n=1}^{\infty}\frac{1}{n^2}
    \exp\!\left(-\frac{n\pi m^2}{|\mathfrak qE|}\right) \quad \text{(non-perturbative, all orders resummed)},
\label{eq:static_schwinger}
\end{equation}
which is non-perturbative in the external field and resums arbitrarily high orders in $\mathfrak qE$. In the massless limit, $m\to0$, the exponential suppression disappears and we may use $\sum_{n=1}^{\infty}n^{-2}=\pi^2/6$ to find
\begin{equation}
    \left.2{\rm Im}\Gamma_{\rm eff}^{\textcolor{white}{(2)}}
    \right|_{\rm static}
    =
    \frac12\left. {\rm Im}\Gamma_{\rm eff}^{(2)} \right|_{\text{ stochastic};\, \bk\rightarrow0} \quad (m=0).
\label{eq:static_stochastic_relation}
\end{equation}

This factor of two should not be interpreted as an inconsistency. The soft limit of the stochastic result and the strictly static Schwinger problem are distinct physical limits, and there is no reason for their coefficients to coincide. The stochastic calculation starts from a time-dependent, generally finite-coherence background described by $F_{\mu\nu}(\tau,\bk)$ and evaluates the response perturbatively to quadratic order in the prescribed field. Taking $\bk\rightarrow0$ subsequently probes the long-wavelength sector of this stochastic result, but does not turn the underlying calculation into the non-perturbative constant-field problem.

A strictly static and homogeneous electric field, on the other hand, corresponds to imposing the singular zero mode $k^\mu=0$ from the outset and solving the pair-production problem non-perturbatively in $\mathfrak qE$. The operations of taking the soft limit of the stochastic quadratic kernel and imposing an exactly constant background therefore need not commute,
\begin{equation}
    \lim_{k^\mu\rightarrow0}\Gamma_{\rm eff}^{(2)}[F(k)]
    \neq
    \Gamma_{\rm eff}[F_{\rm const}] .
\label{eq:noncommuting_static_limit}
\end{equation}
In the massless theory the difference is the factor of two in \cref{eq:static_stochastic_relation}. For $m\neq0$, however, the comparison is less direct. Already in flat spacetime the quadratic kernel develops the pair-production threshold 
\begin{equation}
    \sqrt{-K^2}\ \geq\ 2m,
\label{eq:pair_threshold}
\end{equation}
while in de Sitter the corresponding massive stochastic kernel is considerably more involved: massive QED is not conformally invariant, the fermion modes are no longer plane waves, and the kernel acquires genuine dependence on $m/H$ and on the spacetime curvature. Consequently, the simple massless comparison between the soft stochastic limit and the constant-field result does not carry over in any straightforward way. A proper comparison therefore requires the full massive de Sitter noise kernel, which we do not compute here. We leave the detailed analysis of the massive stochastic regime and its relation to the non-perturbative constant-field Schwinger result for future work.

The quadratic stochastic result should therefore not be expected to reproduce the massive constant-field Schwinger effect. Conversely, away from the strict soft limit, it retains information about the temporal and spatial spectrum of the stochastic field and describes pair production by a genuinely time-dependent and inhomogeneous background. The comparison with the static result should therefore be regarded as a useful characterisation of the limiting behaviour of the stochastic kernel, rather than as a matching condition between the two descriptions.

\section{Further applications of the stochastic formalism}
\label{sec:extensions}
The formalism developed above extends beyond the Abelian setup in de Sitter and spatially flat FLRW spacetimes analysed in \cref{sec:particle_creation}. In this section, we generalise the stochastic Schwinger mechanism to weakly coupled non-Abelian gauge sectors, conformally coupled scalar fields in cosmology, and transient sources in asymptotically flat spacetimes.

\subsection{Perturbative pair production by stochastic non-Abelian gauge fields}
\label{sec:chromo}

At low energies, non-Abelian gauge sectors enter qualitatively different infrared regimes. For example, QCD becomes confining below $\Lambda_{\rm QCD}$, while the electroweak $SU(2)_L$ sector is spontaneously broken below the electroweak scale, giving masses to the $W^\pm$ and $Z$ bosons. More generally, dark non-Abelian sectors may likewise confine or undergo symmetry breaking at $\Lambda_{\rm G}$. The situation can be qualitatively different in the early Universe, where the characteristic energy scale may lie well above these infrared scales and the corresponding gauge sector can instead admit a weakly coupled perturbative description \cite{Kolb:1990vq} (see also \cref{fig:cosmic-history}).  In this regime, stochastic non-Abelian gauge backgrounds can be treated perturbatively, providing a natural setting in which to extend the stochastic Schwinger mechanism beyond the Abelian case. The present computation applies whenever the relevant non-Abelian sector remains weakly coupled and thermal corrections can be neglected. This includes non-thermal stages of the early Universe, such as inflation and preheating, as well as situations in which an ambient thermal plasma is present but the gauge sector of interest has not yet thermalised with it, see \cref{fig:regime-validity} and \cref{tab:thermal_history}.

\begin{table}[t]
\centering
\renewcommand{\arraystretch}{1.35}
\begin{tabular}{|
>{\centering\arraybackslash}m{2.6cm}|
>{\centering\arraybackslash}m{3.5cm}|
>{\centering\arraybackslash}m{4.2cm}|
>{\centering\arraybackslash}m{4.3cm}|}
\hline
\textbf{Cosmic era} &
\textbf{Thermal state of the cosmic fluid} &
\textbf{Typical status of the matter sector} &
\textbf{Thermal effects on the matter sector}
\\
\hline

Inflation &
Cold &
Non-thermal &
None
\\
\hline

Preheating &
Highly occupied, far from equilibrium &
Non-thermal &
Equilibrium thermal effects negligible
\\
\hline

Reheating &
Thermal bath building up &
Model dependent; some sectors may still satisfy $\Gamma_{\rm th}\ll H$ &
Negligible if the relevant sector is not yet thermalised
\\
\hline

Radiation era &
Hot thermal plasma &
SM typically thermalised; weakly coupled hidden sectors may remain decoupled &
Important for thermalised sectors; negligible if $\Gamma_{\rm th}\ll H$ or $\Lambda_{\rm process}\gg T,m_{\rm th}$
\\
\hline

Matter/dark-energy eras &
No universal hot thermal plasma &
Most cosmological relic sectors are decoupled &
Negligible on cosmic scales, except for processes occurring within local thermal environments
\\
\hline

\end{tabular}
\caption{Thermal history of the Universe and the regimes in which thermal effects on a given matter sector may be neglected.}
\label{tab:thermal_history}
\end{table}

We consider a massless fermion transforming in a representation $\bR$ of a local $SU(N)$ gauge symmetry, described by the action
\begin{equation}
    S
    =
    \int d^4x\sqrt{-g}\ \left[-\frac{1}{4}F_{\mu\nu}^aF^{\mu\nu\, a}+\bar\Psi\!\!\left(i\slashed{\nabla}-\mathfrak{g}\gamma^\mu A_\mu^a\,T^a_{\bR}\right)\!\Psi\right],
\end{equation}
where the index $a$ labels each of the $(N^2-1)$ gauge bosons, $\mathfrak{g}$ is the charge of the fermion, and $T_{\bR}^a$ are the generators of $SU(N)$ in the representation $\bR$.  The coupling between gauge and matter degrees of freedom is given by
\begin{equation}
    \sqrt{-g}\,\mathcal{L}_{\text{int}}=\mathfrak{g}\,\bar{\psi}\gamma^\mu  A_\mu^a T_{\bR}^a \psi,
\end{equation}
where we have used classical conformal invariance to map the fermion degrees of freedom to Minkowski space. At the high energy scales relevant to the early Universe, where the characteristic scale lies well above any confinement or symmetry-breaking scale and the corresponding non-Abelian coupling is weak, a perturbative treatment of the type developed below is justified. For example, this corresponds schematically to $\Lambda\gg\Lambda_{\rm QCD}\sim 200 \,{\rm MeV}$ for QCD and $\Lambda\gg \Lambda_{\rm EW}\sim 100\, {\rm GeV}$ for the electroweak $SU(2)_L$ sector.  The self-interacting nature of the gauge bosons nonetheless means that, depending on the matter content and on the process under consideration, the background may produce either fermion pairs,
\begin{equation}
    \bar{A}^a_\mu
    \longrightarrow
    \psi+\bar\psi ,
\end{equation}
or non-Abelian gauge bosons,
\begin{equation}
    \bar{A}_\mu^a
    \longrightarrow
    A_\mu^b+A_\nu^c
\end{equation}
where $\bar{A}_\mu^a$ denotes a classical background configuration. In what follows we focus on the production of on-shell fermions.  In \cref{tab:nonabelian-extensions}, we compile SM and BSM examples of this process. Leading-order fermion-pair production proceeds through the fermion loop of \cref{fig:Feynman}, with ghost vertices and self-interactions entering at higher order.
\begin{table}[ht]
    \centering
    \begin{tabular}{|c|c|c|c|c|}
        \hline
        &
        Gauge sector
        &
        Relevant regime
        &
        Produced fermions
        &
        Charge replacement
        \\
        \hline
        \multirow{2}{*}{SM}
        &
        $SU(3)_{\rm c}$
        &
        $\Lambda>\Lambda_{\rm QCD}$
        &
        $q+\bar q$
        &
        $
        \mathfrak{q}^2
        \mapsto
        \mathfrak{g}_c^2\,T({\mathbf{3}}_c)
        $
        \\
        %\cline{2-5}
        &
        $SU(2)_L$
        &
        $\Lambda>\Lambda_{\rm EW}$
        &
        $q_L+\bar q_L,\,
        \ell_L+\bar\ell_L$
        &
        $
        \mathfrak{q}^2
        \mapsto
        \tfrac{1}{2}\mathfrak{g}_L^2\,T({\mathbf{2}}_L)
        $
        \\
        \hline
        BSM
        &
        $G_d$
        &
        $\Lambda>\Lambda_{\rm G}$
        &
        $X+\bar X$
        &
        $
        \mathfrak{q}^2
        \mapsto
        \mathfrak{g}_d^2\,T({\mathbf{R}}_d)
        $
        \\
        \hline
    \end{tabular}
    \caption{
    Examples of the perturbative extension of the Abelian stochastic pair-production formalism to non-Abelian gauge sectors.  For QCD-like backgrounds above the confinement scale $\Lambda_{\rm QCD}$, the stochastic gauge field produces quark--antiquark pairs. Above the electroweak phase transition $\Lambda_{\rm EW}$, an $SU(2)_L$ background produces left-handed quark and lepton doublets; because only one Weyl component of the massless Dirac fermion is charged under $SU(2)_L$, its charge replacement carries an additional factor of $\tfrac12$. As a BSM example, we consider a compact Lie group $G_d$ under which the fermionic matter $X$ transforms in a unitary representation $\bR_d$ and chosen such that the theory is free of gauge anomalies.
    }
    \label{tab:nonabelian-extensions}
\end{table}

\begin{figure}[ht]
    \centering
    \includegraphics[width=0.9\linewidth,trim=8cm 10cm 8cm 12cm,clip]{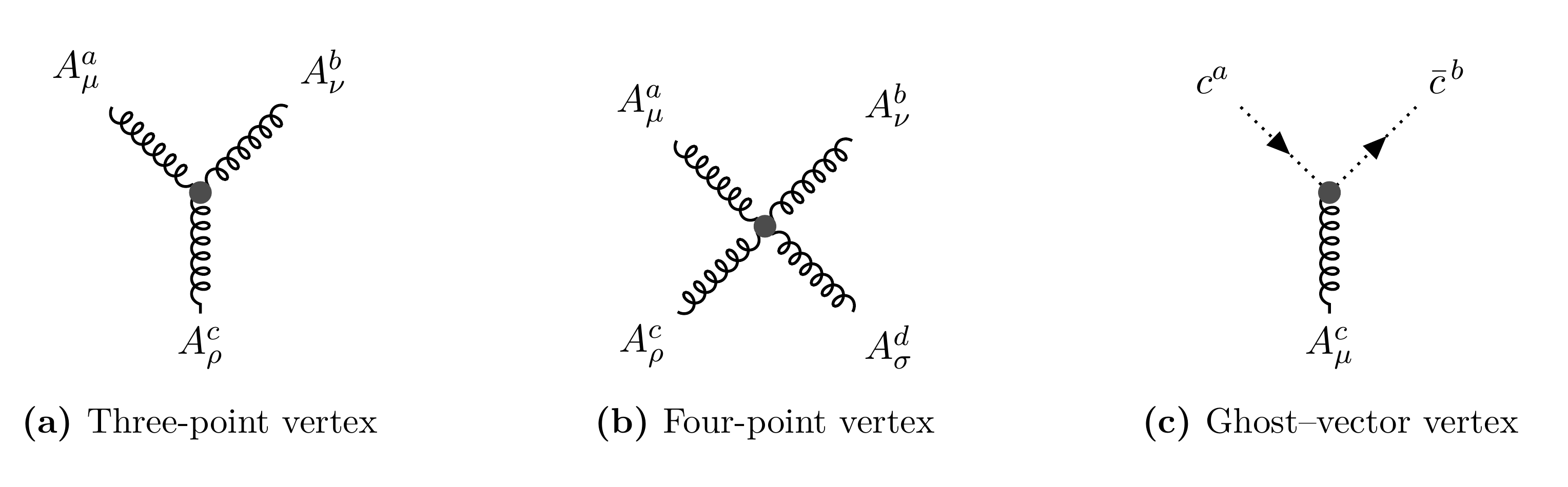}
    \caption{Yang-Mills interaction vertices}
    \label{fig:vertices}
\end{figure}
Ghost and vector loops nevertheless become important once quantum corrections to the gauge-field correlator are included. Schematically, these corrections modify $\langle A_\mu^a A_\nu^b\rangle_{\rm st}$ through the vertices in \cref{fig:vertices}. They are required for a gauge-consistent treatment of the quantum Yang--Mills sector, including vacuum polarisation, coupling renormalisation, and the dynamical evolution of the stochastic gauge background. In the present work, however, we restrict the analysis to the dominant contribution in perturbation theory. At this order, fermion production is governed by the same kinematic kernel as in the Abelian case, supplemented by the appropriate non-Abelian group-theory factor; genuinely non-Abelian effects arise beyond this leading approximation and lie outside the scope of the present analysis.

Expanding the fermion determinant to quadratic order in the background field gives
\begin{equation}
\text{Im}\Gamma_{\rm eff}[A^\mathbf{a}_\mu]=\frac{1}{2}
    \int d^4x\,d^4y\;\langle A^{a,\mathbf{a}}_\mu(x)\,N^{\mu\nu}_{\, ab}(x,y)\,A^{b,\mathbf{a}}_\nu(y)\rangle_{\text{st}}+\mathcal{O}(\mathfrak{g}^4),
\end{equation}
where
\begin{equation}
    N^{\mu\nu}_{\, ab}(x,y)=\mathfrak{g}^2\,\Tr_{\bR}[T_{\bR}^aT_{\bR}^b]\,N^{\mu\nu}(x,y)=\mathfrak{g}^2\, T(\mathbf{R})\, \delta_{ab}\, N^{\mu\nu}(x,y),
\end{equation}
with $T(\bR)$ denoting the Dynkin index of the fermionic representation $\bR$, and $N^{\mu\nu}$ the Abelian noise kernel. The effective action at quadratic order is then
\begin{equation}
    \mathrm{Im}\Gamma_{\rm eff}
    =
    \frac{\mathfrak{g}^2\pi T({\bR})}{6}
    \int d\tau d\tau'\!\int d^3\bk\ 
  \mathcal{I}_k(\tau-\tau') \, 
    \langle 
    -\tfrac{1}{2}F^{\mathbf{a},a}_{\mu\nu}(\tau,\bk)
    F^{\mu\nu\,{\mathbf{a},a}}(\tau',-\bk)
    \rangle_{\text{st}} +\mathcal{O}(\mathfrak{g}^4).
\end{equation}
Therefore, in the weak-coupling limit of a non-Abelian gauge theory, the result takes the same form as in \cref{eq:ImGamma_final}, with the Abelian charge factor $\mathfrak q^2$ replaced by the corresponding group-theory factor $\mathfrak g^2 T(\bR)$.

\subsection{Massless conformally coupled scalar}\label{sec:conf-scalar}

Conformal invariance under Weyl transformations is not special to spin-$1/2$ fields; the same mapping applies to a massless, conformally coupled charged scalar. Such fields arise naturally in four-dimensional conformal gauge theories, for instance in the scalar sector of $\mathcal N=4$ super-Yang--Mills, where the scalars are exactly massless and conformally coupled, with $\xi=1/6$ \cite{Pestun:2007rz}.
Consider a complex scalar $\Phi$ of charge $\mathfrak{q}$, with action
\begin{equation}
    S_{\text{sQED}}
    =
    \int d^4x\sqrt{-g}\ \left[-\frac{1}{4}F_{\mu\nu}F^{\mu\nu}-g^{\mu\nu}(D_\mu\Phi)^\dagger D_\nu\Phi-\left(m^2+\varsigma R\right)|\Phi|^2\right],
\label{eq:conformal-scalar}
\end{equation}
where $D_\mu=\partial_\mu+i \mathfrak{q} A_\mu$, $\varsigma$ is the non-minimal coupling parameter, and $R$ is the Ricci curvature. For the conformal values,
\begin{equation}
    m=0\quad \text{and}\quad \varsigma=\frac{1}{6},
\end{equation}
the action is invariant under the same Weyl transformations as in \cref{sec:conf}, now with
\begin{equation}
    \Phi(x)\mapsto\Omega^{-1}(x)\Phi(x).
\end{equation}
Choosing $\Omega(x)=a^{-1}(\tau)$ as before and introducing the conformally rescaled field
\begin{equation}
    \varphi(x)\equiv a(\tau)\,\Phi(x),
\end{equation}
the action maps onto its flat-space counterpart,
\begin{equation}
    S_{\text{sQED}} \mapsto \int d^4x\left[-\frac14F_{\mu\nu}F^{\mu\nu}-\eta^{\mu\nu}(D_\mu\varphi)^\dagger D_\nu\varphi\right].
\end{equation}
The rescaled field $\varphi$ therefore obeys the free Klein--Gordon equation of ordinary flat-space scalar electrodynamics, and cosmological expansion alone cannot excite this sector. As for the massless fermion, any pair production of conformally coupled scalars must be driven by the external stochastic gauge background rather than by the geometry itself \cite{Parker:2009uva}.

The only modification at leading order relative to \cref{sec:particle_creation} is the matter-loop coefficient: a complex conformally coupled scalar contributes a non-local vacuum-polarisation piece equal to $1/4$ that of a single Dirac fermion of the same charge, so that
\begin{equation}
    \mathrm{Im}\Gamma_{\rm eff}^{\rm sQED}
    =
    \frac14\,
    \mathrm{Im}\Gamma_{\rm eff}
    =
    \frac{\mathfrak{q}^2\pi}{24}
    \int d\tau\,d\tau'\!\int d^3\bk\ 
    \mathcal I_k(\tau-\tau')
    \left\langle
    -\tfrac12\,
    F_{\mu\nu}(\tau,\bk)
    F^{\mu\nu}(\tau',-\bk)
    \right\rangle_{\rm st}
    +\mathcal{O}(\mathfrak{q}^3).
\end{equation}
Possible local contributions associated with renormalisation and the conformal anomaly are distinct from the non-local imaginary part considered here and do not modify the kernel $\mathcal I_k$. The generalisation to a non-Abelian gauge group proceeds exactly as in \cref{sec:chromo} for the fermionic case, with the scalar loop coefficient above supplemented by the same factor of $T(\bR)$.

\subsection{Non-stationary  backgrounds in asymptotically flat spacetimes}
\label{subsec:nonstationary-flat}

A stochastic gauge-field background need not be cosmological in origin to be genuinely non-stationary. Relativistic magnetic reconnection provides a natural astrophysical example: although the ambient plasma may be magnetically dominated, the reconnecting magnetic field is strongly suppressed near current sheets and X-points while the reconnection electric field remains finite, producing localised electric-dominated regions with $E^2>B^2$ \cite{Uzdensky:2011,Sironi:2022}. In turbulent or plasmoid-dominated reconnection, such regions are created and destroyed intermittently, so their amplitudes, locations, and lifetimes are naturally described statistically. A laboratory analogue is provided by colliding intense laser pulses, for which the magnetic fields can cancel near electric antinodes while the electric fields add, yielding transient regions with $E^2>B^2$; fluctuations in pulse timing, phase, or amplitude can further render the background stochastic \cite{Kohlfurst:2021,Sah:2025}. A related BSM example is an ultralight dark-photon background, for which the dark magnetic field is velocity suppressed, $B_D\sim v E_D$, so that $E_D^2>B_D^2$ for a non-relativistic field, while finite coherence times and random phases provide a direct statistical description \cite{Graham:2020,Centers:2019}. In all these cases the ambient spacetime is flat to excellent approximation, while the relevant gauge background is transient, statistically varying, and electric dominated.  Such systems naturally contain a coherent component together with rapidly evolving fluctuations. Writing $F_{\mu\nu}=\overline F_{\mu\nu}+\delta F_{\mu\nu}$, with $\langle\delta F_{\mu\nu}\rangle_{\rm st}=0$, we focus here on the stochastic contribution $\delta F_{\mu\nu}$ and henceforth denote it simply by $F_{\mu\nu}$. The present perturbative treatment applies when this fluctuating component can be treated at $\mathcal{O}(\mathfrak q^2)$ without resumming the coherent background into the fermion propagator.

The distinction between stationary and non-stationary stochastic sources is particularly transparent in the mixed $(\tau,\bk)$ representation used throughout this work. Assuming spatial statistical homogeneity, a stationary source satisfies
\begin{equation}
    \left\langle\delta F_{\mu\nu}(k^{\mu}) \delta F^{\mu\nu}(-k'^{\mu})\right\rangle_{\rm st}
    \propto
    \delta^{(4)}(k^\mu-k'^\mu) \quad \text{(stationary)}.
\end{equation}
A genuinely transient source does not possess this statistical time-translation invariance 
\begin{equation}
    \left\langle\delta F_{\mu\nu}(\tau,\bk)\delta F^{\mu\nu}(\tau',\bk') \right\rangle_{\rm st}
    =
    \delta^{(3)}(\bk-\bk') \, \mathcal P_{\bk}(\tau,\tau')  \quad \text{(non-stationary)},
\end{equation}
Thus, although a double Fourier transform may always be introduced formally, there is in general no frequency-diagonal stationary spectrum and no unique frequency characterising the source. This is precisely the regime in which retaining the explicit two-time dependence is advantageous. If the statistics evolve slowly compared with the correlation time, a short-time Fourier transform can instead provide an approximately local spectral description, as considered in our original flat-space analysis \cite{VicenteGarcia-Consuegra:2025lkh}. Finite-duration electric pulses and, more recently, pulse trains with stochastic time delays provide related $U(1)$ examples in which the temporal structure and randomness of the source directly affect pair production \cite{Sah:2025}.

Importantly, the absence of time-translation invariance in the \emph{source} does not remove the time-translation invariance of the free Minkowski matter sector. The fermionic two-point function, and hence the quadratic absorptive kernel, therefore continues to depend only on $|\tau-\tau'|$. Equivalently, setting $a(\tau)=1$ in the conformally flat analysis of \cref{sec:particle_creation} gives the same kernel $\mathcal I_k(\tau-\tau')$, while the stochastic source retains its full two-time dependence. For one massless Dirac fermion, the resulting quadratic vacuum-decay exponent is
\begin{equation}
    2{\rm Im}\Gamma_{\rm eff}
    =
    \frac{\mathfrak q^2\pi}{3}\int d\tau \,d\tau' \int d^3\bk\ \mathcal I_k(\tau-\tau')\left\langle-\tfrac12F_{\mu\nu}(\tau,\bk)F^{\mu\nu}(\tau',-\bk)\right\rangle_{\rm st}+\mathcal O(\mathfrak q^4).
\end{equation}
This expression therefore requires neither stationarity nor the assignment of a definite frequency to the external field: all information about the non-stationary source is retained in its unequal-time field correlator.
Asymptotically flat spacetimes moreover admits genuine asymptotic particle states whenever the interaction switches off sufficiently rapidly in the far past and future,
\begin{equation}
    A_\mu(\tau\rightarrow\pm\infty,\bx)\longrightarrow0.
\end{equation}
For further details on asymptotic gauge-field fall-off conditions and related memory effects, see \cite{Strominger:2013lka,Strominger:2017zoo,Maleknejad:2023nwe}.
For each realisation, the fermion field then approaches the free theory asymptotically, allowing unambiguous in- and out-vacua and a Bogoliubov transformation between the corresponding mode bases.

\section{Conclusions and future directions\label{sec:concl}}

In this work, we developed a stochastic formulation of the Schwinger effect in de Sitter spacetime using the Schwinger--Keldysh (in-in) formalism. Treating the gauge field as a prescribed classical stochastic background, we integrated out the quantum matter sector and derived the corresponding influence functional and particle-production kernel to leading non-trivial order in the gauge coupling. For massless charged matter, conformal symmetry provides a particularly powerful simplification: the matter sector in spatially flat FLRW spacetimes can be mapped to its flat-space counterpart, allowing the cosmological calculation to remain analytically tractable while retaining the full time dependence and stochasticity of the gauge background.

In \cref{sec:particle_creation}, we derived the fermionic CTP influence functional and identified its imaginary part with the stochastic matter-production probability (see \cref{eq:ImGamma_final}). Quantising the massless Dirac field in de Sitter and evaluating the corresponding noise kernel, we obtained the production rate directly in the mixed representation $F_{\mu\nu}(\tau,\bk)$, without introducing asymptotic out states or assuming a stationary background. Owing to conformal invariance, the resulting kernel is independent of the scale factor, allowing the de Sitter result to extend directly to generic spatially flat FLRW spacetimes (see \cref{sec:generic_FLRW}). 

We also clarified the consistency and limiting regimes of the stochastic description. For massless fermions in de Sitter, conformal invariance prevents an intrinsic infrared enhancement of the matter correlators, so the long-wavelength behaviour remains regular provided that the prescribed stochastic gauge spectrum is itself infrared safe (see \cref{sec:IR}). The gauge background must moreover be sufficiently highly occupied to admit a classical stochastic description while remaining subdominant in the total cosmic energy density (see \cref{eq:class-BG}). Finally, the soft limit of the perturbative stochastic result should be distinguished from the strictly constant-field Schwinger problem: the former is obtained from the quadratic response to a time-dependent stochastic background, whereas the latter is intrinsically non-perturbative and resums the external field to all orders. One may formally consider the zero-momentum (long-wavelength and low-frequency) limit of our stochastic result in 4D de sitter, while keeping the electric-field amplitude finite. This gives $\Upsilon_{\rm eff}/V_4\to \mathfrak q^2E^2/(12\pi)$, a factor of two larger than the exact massless static Schwinger result, $\mathfrak q^2E^2/(24\pi)$. The discrepancy reflects the breakdown of the quadratic perturbative expansion in this limit: at fixed $E$, $\bk\rightarrow0$ drives the system into the non-perturbative constant-background regime, which requires an all-order resummation \cite{Ferreiro:2023jfs} (see \cref{sec:constant_field_limit} for details).
As a consistency check, in the flat-space limit $a(\tau)\to1$ and for backgrounds admitting a well-defined frequency representation, the mixed form $F_{\mu\nu}(\tau,\bk)$ reduces to the standard four-momentum description $F_{\mu\nu}(k^\mu)$. In this frequency-resolved limit, our stochastic result reproduces the flat-space expression obtained in \cite{VicenteGarcia-Consuegra:2025lkh}, while retaining a more general formulation for finite-duration and non-stationary backgrounds (see \cref{sec:frequency_resolved_limit}).

Beyond Abelian fermion production in de Sitter, the framework admits several extensions. We extended the perturbative formalism to weakly coupled non-Abelian gauge sectors, with applications to high-energy Standard Model and BSM gauge theories whenever the characteristic scale lies above the relevant confinement or symmetry-breaking scale and thermal corrections are negligible (see \cref{sec:chromo}). We further showed that conformal invariance carries the result directly to spatially flat FLRW backgrounds and to massless conformally coupled scalar matter (see \cref{sec:conf-scalar}).  The same stochastic viewpoint also applies to finite-duration, non-stationary sources in asymptotically flat spacetime (see \cref{subsec:nonstationary-flat}). These extensions illustrate the broad range of settings in which stochastic gauge backgrounds can act as efficient sources of quantum matter creation.

Several directions naturally follow from this work. A central theoretical challenge is to extend the present analytical treatment beyond massless conformal matter, in particular to massive fermions and more general charged fields in de Sitter, where the simple Weyl mapping is lost and genuinely curvature-dependent effects enter the production kernel. It will also be important to move beyond the perturbative regime through dedicated numerical studies, including lattice simulations where appropriate, in order to access strongly time-dependent, highly occupied, and eventually non-perturbative gauge-field dynamics. Another important direction is to embed the stochastic Schwinger mechanism in well-motivated early-Universe scenarios, including inflationary gauge-field production, preheating, and weakly coupled hidden-sector dynamics. In these settings, one can quantify not only the efficiency of matter production, but also its impact on the stability and lifetime of the gauge background, the onset of backreaction, and the redistribution of energy between the gauge and matter sectors. Ultimately, combining improved analytical control with fully dynamical numerical evolution should clarify when particle production remains a spectator effect and when it qualitatively modifies the cosmological background itself.

\section*{Acknowledgements}
We wish to thank Maria Mylova and Cliff P. Burgess for insightful discussions on this topic. LVGC is grateful for the hospitality of the Kavli Institute for the Physics and Mathematics of the Universe at the University of Tokyo, where this work was finalised. AM thanks Géraldine Servant and the DESY Theory Group in Hamburg for their hospitality, where much of this work was carried out with support from the Deutsche Forschungsgemeinschaft under Germany’s Excellence Strategy (EXC 2121 Quantum Universe – 390833306). AM is also grateful to Eiichiro Komatsu, the Max Planck Institute for Astrophysics, and the Munich Institute for Astro-, Particle and BioPhysics (MIAPbP) for their hospitality during the final stages of this work.
 LVGC and AM are supported by the Royal Society under the University Research Fellowship, Grant No. RE22432.

\appendix
\crefalias{section}{appendix}

\section{Dirac fermions in curved spacetimes}
\label{app:Dirac}
This appendix collects the spinor conventions, gamma-matrix identities, and Wightman functions used throughout the main text.
\subsection{Clifford algebra and trace identities}
In flat spacetime, the Dirac gamma matrices satisfy the Clifford algebra, $\text{Cl}_{3,1}$,
\begin{equation}
    \{\gamma^\alpha,\gamma^\beta\}=2\eta^{\alpha\beta}\mathbbm{1}_{4},
\end{equation}
where $\mathbbm{1}_{n}$ is the identity matrix of rank $n$. A convenient explicit representation compatible with this convention is
\begin{equation}
    \gamma^0=
    \begin{pmatrix}
        0 & \mathbbm{1}_{2} \\
        -\mathbbm{1}_{2} & 0
    \end{pmatrix},
    \qquad
    \gamma^i=
    \begin{pmatrix}
        0 & \sigma^i \\
        \sigma^i & 0
    \end{pmatrix},
    \qquad i=1,2,3,
\end{equation}
where $\sigma^i$ are the Pauli matrices. In particular,
\begin{equation}
    (\gamma^0)^2=-\mathbbm{1}_{4},
    \quad\text{and}\quad
    (\gamma^i)^2=\mathbbm{1}_{4}.
\end{equation}
The traces of an odd number of gamma matrices vanish. The trace
identities required in the main text are
\begin{equation}
    \operatorname{tr}\,[\gamma^\alpha\gamma^\beta]=4\eta^{\alpha\beta},
\label{eq:tr-1}
\end{equation}
and
\begin{equation}
    \operatorname{tr}
    \,[ \gamma^\alpha\gamma^\beta \gamma^\delta\gamma^\sigma]
    =
    4\left(\eta^{\alpha\beta}\eta^{\delta\sigma}-\eta^{\alpha\delta}\eta^{\beta\sigma}+\eta^{\alpha\sigma}\eta^{\beta\delta}\right).
\label{eq:tr-2}
\end{equation}
These identities follow directly from the Clifford algebra and are
independent of the particular representation chosen for the gamma
matrices.

\subsection{Dirac equation in curved spacetimes}

In curved spacetime, the position-dependent gamma matrices are defined
in terms of the vierbein $e_\mu{}^\alpha(x)$ and its inverse
$e^\mu{}_\alpha(x)$ as
\begin{equation}
    \gamma^\mu(x)=e^\mu{}_\alpha(x)\gamma^\alpha,
    \quad\text{and}\quad
    \gamma_\mu(x)=e_\mu{}^\alpha(x)\gamma_\alpha,
\end{equation}
where
\begin{equation}
    g_{\mu\nu}(x)=e_\mu{}^\alpha(x)e_\nu{}^\beta(x)\eta_{\alpha\beta},
    \quad\text{and}\quad
    g^{\mu\nu}(x)=e^\mu{}_\alpha(x)e^\nu{}_\beta(x)\eta^{\alpha\beta}.
\end{equation}
Consequently, the curved-spacetime gamma matrices satisfy
\begin{equation}
    \left\{\gamma^\mu(x),\gamma^\nu(x)\right\}
    =
    2g^{\mu\nu}(x)\mathbbm{1}_{4}.
\end{equation}
Their local trace identities are therefore
\begin{equation}
    \operatorname{tr}\ [\gamma^\mu(x)\gamma^\nu(x)]
    =
    4g^{\mu\nu}(x),
\end{equation}
and
\begin{align}
    \operatorname{tr}[\gamma^\mu(x)\gamma^\nu(x)\gamma^\rho(x)\gamma^\lambda(x)]
    =
    4\Big[&g^{\mu\nu}(x)g^{\rho\lambda}(x)-g^{\mu\rho}(x)g^{\nu\lambda}(x)+g^{\mu\lambda}(x)g^{\nu\rho}(x)\Big].
\end{align}

The spinor covariant derivative is
\begin{equation}
    \nabla_\mu\psi=\left(\partial_\mu+\Omega_\mu\right)\psi,
    \quad\text{with}\quad
    \Omega_\mu=\frac{1}{4}\omega_{\mu\alpha\beta}\gamma^{\alpha\beta},
\end{equation}
where
\begin{equation}
    \gamma^{\alpha\beta}
    \equiv
    \frac{1}{2}
    [\gamma^\alpha,\gamma^\beta].
\end{equation}
The curved-spacetime gamma matrices are covariantly constant,
\begin{equation}
    \nabla_\mu\gamma^\nu
    \equiv
    \partial_\mu\gamma^\nu+\Gamma^\nu_{\mu\rho}\gamma^\rho+[\Omega_\mu,\gamma^\nu]=0.
\end{equation}

\subsection{Fermion propagators}\label{app:Wightman}
The greater Wightman function is defined as
\begin{align}
S^>(\tau,\tau';\bk)
&=
\sum_{s=\pm}
U_{s,\bk}(\tau)\bar U_{s,\bk}(\tau')
\nonumber\\
&=
\frac{1}{2}\sum_{s=\pm}
\begin{pmatrix}
u^{\uparrow}_s(\tau,k)\bar u^{\uparrow}_s(\tau',k)\,\bI
&
-u^{\uparrow}_s(\tau,k)\bar u^{\downarrow}_s(\tau',k)
(\boldsymbol{\sigma}\!\cdot\!\hat{\bk})
\\[6pt]
u^{\downarrow}_s(\tau,k)\bar u^{\uparrow}_s(\tau',k)
(\boldsymbol{\sigma}\!\cdot\!\hat{\bk})
&
-u^{\downarrow}_s(\tau,k)\bar u^{\downarrow}_s(\tau',k)\,\bI
\end{pmatrix},
\label{eq:greaterWt}
\end{align}
where $\bI$ is the $2\times2$ identity matrix and $\boldsymbol{\sigma}$ the vector of Pauli matrices. Substituting the Bunch--Davies mode functions \cref{eq:BD} gives
\begin{equation}
\label{eq:GreaterW_modesapp}
    S^{>}(\tau,\tau';\bk)
    =\frac{1}{2k} \, \frac{1}{(2\pi)^3}\left(\gamma^0k-\boldsymbol{\gamma}\!\cdot\!{\bk}\right)e^{-ik(\tau-\tau')}
    = -\frac{\gamma^\mu{K}_\mu}{2k} \, \frac{e^{-ik(\tau-\tau')}}{(2\pi)^3} ,
\end{equation}
with ${K}^\mu=(k,\bk)$ an auxiliary null four-vector and $K_\mu=(-k,\bk)$. The lesser Wightman function, $S^{<}(\tau,\tau';\bk)= -\sum_{s=\pm}V_{s,\bk}(\tau)\bar V_{s,\bk}(\tau')$, evaluates analogously using the charge-conjugation relations \cref{eq:uvpl},
\begin{equation}
\label{eq:LesserW_modes}
    S^{<}(\tau,\tau';\bk)
    = -\frac{1}{2k}\, \frac{1}{(2\pi)^3}\left(\gamma^0k-\boldsymbol{\gamma}\!\cdot\!{\bk}\right)e^{ik(\tau-\tau')}
    =\frac{\gamma^\mu{K}_\mu}{2k} \frac{e^{ik(\tau-\tau')}}{(2\pi)^3}.
\end{equation}
The two satisfy the standard adjoint relation
\begin{equation}
\label{eq:Wightman_adjoint_relation}
    S^<(\tau,\tau';\mathbf k)
    = { -S^>(\tau',\tau;\mathbf k) = -\gamma^0\left[S^>(\tau,\tau';\mathbf k)\right]^\dagger\gamma^0 },
\end{equation}
so that \cref{eq:GreaterW_modesapp,eq:LesserW_modes} are the fermionic propagators entering the one-loop noise kernel.

\section{Technical details of the imaginary part of the effective action}
\label{app:deets}

In this appendix, we provide details of the calculation of the imaginary part of the QED effective action.

\subsection{Momentum structure}\label{App:momentum_structure}

Here we derive the imaginary part of the influence functional $S_{\rm IF}$ in momentum-space and work out the form presented in \cref{eq:IM_action}. Since the current operator is Hermitian, $(\hat j^\mu)^\dagger=\hat j^\mu$, the Wightman functions satisfy
\begin{equation}
G_{+-}^{\mu\nu}(x,y)
\equiv
\langle \hat j^\nu(y)\hat j^\mu(x)\rangle=
\left[
G_{-+}^{\mu\nu}(x,y)
\right]^*,
\end{equation}
where $G_{-+}^{\mu\nu}(x,y)
\equiv
\langle \hat j^\mu(x)\hat j^\nu(y)\rangle$.
It follows that
\begin{equation}
\langle\{\hat j^\mu(x),\hat j^\nu(y)\}\rangle=
G_{-+}^{\mu\nu}(x,y)
+
G_{+-}^{\mu\nu}(x,y)
=
2\,{\rm Re}\,G_{-+}^{\mu\nu}(x,y),
\end{equation}
whereas
\begin{equation}
\langle[\hat j^\mu(x),\hat j^\nu(y)]\rangle =G_{-+}^{\mu\nu}(x,y)-
 G_{+-}^{\mu\nu}(x,y)
=
2i\, {\rm Im}\,G_{-+}^{\mu\nu}(x,y).
\end{equation}
The anticommutator is therefore real, while the commutator is purely imaginary. Consequently, in the CTP effective action, the symmetrised correlator governs the imaginary noise and decoherence kernel, whereas the commutator determines the real causal-response kernel.

Using the above in the position space \cref{eq:Gamma2_CTP,eq:G-}, we find
\begin{equation}
    \text{Im}\Gamma_{\text{eff}}   [A^{\mathbf{a}}_\mu] = \frac14 \int d\tau\, d\tau' \int d^3\bx \, d^3\by\ \langle\{ \hat{j}^\mu(\tau,\bx),\hat{j}^\nu(\tau',\by)\}\rangle\ \langle A^{\mathbf{a}}_\mu(\tau,\bx) A^{\mathbf{a}}_\nu(\tau',\by)\rangle_{st}.
\end{equation}
Substituting \cref{eq:Pi-N,eq:noise_Wightman} into the expression above, and using \cref{eq:S-gtrless}, we obtain
\begin{align}
    \text{Im}\Gamma_{\text{eff}}
   = & -  \frac{\mathfrak{q}^2}{4} \, \int d\tau\, d\tau'\int d^3\bx \, d^3\by\int d^3\bq_1\, d^3\bq_2\, d^3\bp_1\, d^3\bp_2\;e^{i\bx\cdot(\bq_1-\bp_1-\bp_2)}e^{i\by\cdot(\bq_2+\bp_1+\bp_2)}\nonumber\\
    &\quad\times\mathrm{tr}\!\left[\gamma^\mu S^>(\tau,\tau';\bp_1)\gamma^\nu S^<(\tau',\tau;\bp_2)\right]\langle A^{\mathbf{a}}_\mu(\tau,\bq_1) A^{\mathbf{a}}_\nu(\tau',\bq_2)\rangle_{st}+\text{h.c.} ,
\end{align}
where $``\text{h.c.}"$ denotes Hermitian conjugation. Performing the integrals over $\bx$, $\by$, $\bq_1$, and $\bq_2$, we arrive at 
\begin{align}
    \text{Im}\Gamma_{\text{eff}}
  & =  -  \frac{\mathfrak{q}^2 \, (2\pi)^6}{4} \, \int d\tau\, d\tau' \int_{\bp_1,\bk} \,  \mathrm{tr}\!\left[\gamma^\mu S^>(\tau,\tau';\bp_1)\gamma^\nu S^<(\tau',\tau;\bk-\bp_1)\right]\langle A^{\mathbf{a}}_\mu(\tau,\bk) A^{\mathbf{a}}_\nu(\tau',-\bk)\rangle_{st} \nonumber\\
   &+\text{h.c.}.
\end{align}
Finally, we can write $\text{Im}\Gamma_{\text{eff}}$ as 
\begin{equation}
    \text{Im}\Gamma_{\text{eff}}[A^\mathbf{a}_\mu]= \frac12\int d\tau\, d\tau'\int d^3\bk\ N^{\mu\nu}_{\bk}(\tau,\tau')\,\langle A^\mathbf{a}_\mu(\tau,\bk) A^\mathbf{a}_\nu(\tau',-\bk)\rangle_{st},
\end{equation}
where
\begin{align}
    N^{\mu\nu}_{\bk}(\tau,\tau')
    \equiv -  \frac{\mathfrak{q}^2 \, (2\pi)^6}{2} \, \int d^3\bp_1\ \mathrm{tr}\!\left[\gamma^\mu S^>(\tau,\tau';\bp_1)\gamma^\nu S^<(\tau',\tau;\bk-\bp_1)\right]+\text{h.c.},
\end{align}
is the momentum-space fermionic noise kernel.

\subsection{Loop integral}
\label{subsec:momentum-integrals}
In this subsection we collect the details of the loop integral appearing in the evaluation of the noise kernel. Using \cref{eq:greaterWt}, we can write the integral encountered in \cref{eq:Nk-munu} as
\begin{align}
    N^{\mu\nu}_{\bk}(\tau,\tau') &
    = \mathfrak{q}^2 \, \int \frac{d^3 \bp_1}{p_1 p_2}\,\left[P_1^\mu P_2^\nu+P_1^\nu P_2^\mu-(P_1\!\cdot\!P_2)\eta^{\mu\nu}\right] \cos[(p_1+p_2)(\tau-\tau')]\big\vert_{\bp_2=\bk-\bp_1},
\label{eq:basic-momentum-integral}
\end{align}
where
\begin{equation}
P_1^0=|\bp_1|,
\qquad
P_2^0=|\bk-\bp_1|.
\end{equation}
 After making use of the spectral decomposition
\begin{equation}
    \cos[(p_1+p_2)(\tau-\tau')]
    =
    \frac12
    \int_{-\infty}^{\infty}
    d\omega\;
    e^{-i\omega(\tau-\tau')}
    \Big[
    \delta(p_1+p_2-\omega)
    +
    \delta(p_1+p_2+\omega)
    \Big],
\end{equation}
we can write the kernel in a spectral four-momentum representation
\begin{align}
    N^{\mu\nu}_{\bk}(\tau,\tau') &
    \equiv \frac{\mathfrak{q}^2}{2} \, \int dK^0 \, e^{-iK^0(\tau-\tau')} \, \mathcal{N}^{\mu\nu}(K),
    \label{eq:N-mu-nu}
\end{align}
where $\mathcal{N}^{\mu\nu}(K)$ is defined as
where
\begin{equation}
\mathcal{N}^{\mu\nu}(K)
=
\mathcal{N}^{\mu\nu}_{+}(K)
+
\mathcal{N}^{\mu\nu}_{-}(K),
\end{equation}
with
\begin{align}
\mathcal{N}^{\mu\nu}_{\pm}(K)
&\equiv 2
\int \frac{d^3\bp_1}{p_1p_2}\,
\left[
P_1^\mu P_2^\nu
+
P_1^\nu P_2^\mu
-
(P_1.P_2)\eta^{\mu\nu}
\right]
\delta(p_1+p_2 \mp K^0).
\end{align}
For the positive-frequency branch, $K^0 > 0$, momentum conservation gives
\begin{equation}
K^\mu=P_1^\mu+P_2^\mu .
\end{equation}
Since the tensor structure obtained below is even under this transformation, the two branches combine into a result depending only on $K^2$, with support in the timelike region $K^2<0$. The negative-frequency branch is obtained by 
\begin{align}
\mathcal{N}^{\mu\nu}_{-}(K^0,\bk)
= \mathcal{N}^{\mu\nu}_{+}(-K^0,\bk).
\end{align}
 It is therefore sufficient to evaluate the positive-frequency branch. At $K^0=0$, the two delta functions coincide and reduce to
$2\delta(p_1+p_2)$. Since the on-shell energies satisfy $p_1,p_2\geq0$, the only support is at the degenerate point $p_1=p_2=0$, which also implies $\bk=\boldsymbol{0}$. There is therefore no non-trivial two-particle phase space at $K^0=0$, and the spectral kernel vanishes at this isolated point,
\begin{equation}
\mathcal N^{\mu\nu}(K^0=0,\bk)=0.
\end{equation}
This should be distinguished from the null boundary $K^2=0$ with $K^0=|\bk|>0$, which corresponds to finite-energy collinear massless momenta. The null result is obtained continuously from the timelike region. We therefore evaluate the phase-space integral for $K^2<0$ in the center-of-momentum frame and extend the resulting expression to the null boundary by continuity.

The kernel is Lorentz covariant and therefore must depend only on the external momentum $K^\mu$ and the metric $\eta^{\mu\nu}$, i.e.
\begin{equation}
    \mathcal{N}^{\mu\nu}(K)
    =\mathcal{A}(K^2)\eta^{\mu\nu}+\mathcal{B}(K^2)K^\mu K^\nu .
\label{eq:I_tensor_ansatz}
\end{equation}
The coefficients may be fixed by taking scalar projections. We note that since the internal propagators are massless fermions and $K^\mu=p^\mu+p'^\mu$, we have
\begin{equation}
    P_1^2=P^2_2=0,
    \quad \text{and}\quad
    P_1\cdot P_2=\frac{K^2}{2}.
\end{equation}

First, we compute the trace of the integral of the positive-frequency branch as
\begin{equation}
    \eta_{\mu\nu}\mathcal{N}_+^{\mu\nu}(K)= -2K^2\int \frac{d^3 \bp_1}{p_1 p_2} \, \delta(p_1+p_2-K^0) = -4\pi K^2 \, \quad \text{for} \quad K^0>0.
\end{equation}
Second, contracting with $K_\mu K_\nu$ gives
\begin{align}
    K_\mu K_\nu \mathcal{N}_+^{\mu\nu}(K)
    &= 2\int \frac{d^3\bp_1}{p_1 p_2}\, K_\mu K_\nu \left[P_1^\mu P_2^\nu +P_1^\nu P_2^\mu -(P_1\cdot P_2)\eta^{\mu\nu}\right]\delta(p_1+p_2-K^0)=0,
\end{align}
as expected from the transversality of the kernel. The projections of our ansatz in  \cref{eq:I_tensor_ansatz} are then
\begin{align}
    \eta_{\mu\nu}\mathcal{N}_+^{\mu\nu}
    &=
    (4\mathcal{A}+\mathcal{B}\,K^2) \, \Theta(K^0)=-4\pi K^2 \, \Theta(K^0),
    \\
    K_\mu K_\nu \mathcal{N}_+^{\mu\nu}
    &=
    \mathcal{A}\,K^2+\mathcal{B}\,(K^2)^2=0.
\end{align}
Solving for $\mathcal{A}$ and $\mathcal{B}$ gives
\begin{equation}
    \mathcal{A}=-\frac{2\pi}{3}K^2,\qquad\mathcal{B}=\frac{2\pi}{3},
\end{equation}
and hence
\begin{equation}
        \mathcal{N}_+^{\mu\nu}(K)
        =\frac{4\pi}{3}\left(
        K^\mu K^\nu-K^2\eta^{\mu\nu}\right)\Theta(-K^2) \, \Theta(K^0).
\label{eq:I_covariant}
\end{equation}
Since the negative-frequency contribution is obtained by reflecting the positive-frequency branch, the two branches combine to give
\begin{equation}
        \mathcal{N}^{\mu\nu}(K)
        =\frac{4\pi}{3}\left(
        K^\mu K^\nu-K^2\eta^{\mu\nu}\right)\Theta(-K^2).
\label{eq:I_covariant_result}
\end{equation}

\subsection{Integration by parts and field-strength representation}
Substituting \cref{eq:I_covariant_result} into \cref{eq:N-mu-nu}, we find 
\begin{align}
    N^{\mu\nu}_{\bk}(\tau,\tau') &
    \equiv \frac{\pi \, \mathfrak{q}^2}{3} \, \int dK^0 \, e^{-iK^0(\tau-\tau')} \, \left(
        K^\mu K^\nu-K^2\eta^{\mu\nu}\right)\Theta(-K^2).
    \label{eq:N-mu-nu--}
\end{align}
 As discussed in the main text, $K^0$ is a spectral variable inherited from the conformally rescaled fermion loop and should not be interpreted as a physical photon frequency. 
The components of the tensor $N_{\bk}^{\mu\nu}$ can all be expressed
in terms of the scalar kernel
\begin{equation}
\label{eq:Ikernelapp}
    \mathcal I_k(\Delta\tau)\equiv \int_k^\infty d\omega\,\left(e^{-i\omega\Delta\tau} +e^{i\omega\Delta\tau}\right)
    =
    2\left[\pi\delta(\Delta\tau)-\frac{\sin(k\Delta\tau)}{\Delta\tau}\right],
    \qquad
    \Delta\tau\equiv\tau-\tau',
\end{equation}
where $k\equiv|\bk|$ and the second equality is understood in the
distributional sense. In terms of this kernel, the individual components
satisfy the following relations:
\begin{enumerate}

\item \textit{Time-time} component:

\begin{equation}
    N_{\bk}^{00}(\tau,\tau')=\frac{\pi \mathfrak{q}^2}{3}\,k^2\mathcal I_k(\tau-\tau').
\end{equation}

\item \textit{Mixed components:}

\begin{enumerate}

\item \textit{Time-space} component:

\begin{align}
    N_{\bk}^{0i}(\tau,\tau')
    &=-\frac{\pi \mathfrak{q}^2 k^i}{3}\int_k^\infty d\omega\,\omega\left[e^{-i\omega(\tau-\tau')}-e^{i\omega(\tau-\tau')}\right]\nonumber\\
    &=\frac{i\pi \mathfrak{q}^2 k^i}{3}\int_k^\infty d\omega\,\partial_{\tau'}\left[e^{-i\omega(\tau-\tau')}+e^{i\omega(\tau-\tau')}\right]
    =\frac{i\pi \mathfrak{q}^2 k^i}{3}\,\partial_{\tau'}\mathcal I_k(\tau-\tau').
\end{align}

\item \textit{Space-time} component:

\begin{align}
    N_{\bk}^{i0}(\tau,\tau')
    &=-\frac{\pi \mathfrak{q}^2 k^i}{3}\int_k^\infty d\omega\,\omega\left[e^{-i\omega(\tau-\tau')}-e^{i\omega(\tau-\tau')}\right]\nonumber\\
    &=-\frac{i\pi \mathfrak{q}^2 k^i}{3}\int_k^\infty d\omega\,\partial_\tau\left[e^{-i\omega(\tau-\tau')}+e^{i\omega(\tau-\tau')}\right]
    =-\frac{i\pi \mathfrak{q}^2 k^i}{3}\,\partial_\tau\mathcal I_k(\tau-\tau').
\end{align}
\end{enumerate}
The tensor structure is manifestly symmetric under $\mu\leftrightarrow\nu$, while time-translation invariance implies that the kernel depends only on the time difference $\Delta\tau\equiv\tau-\tau'$. Therefore, the mixed components satisfy
\begin{equation}
    N^{0i}_{\bk}(\Delta\tau)=N^{i0}_{\bk}(\Delta\tau)=
    \frac{i\pi \mathfrak{q}^2}{3}\,k^i\frac{\partial}{\partial\Delta\tau}\mathcal I_k(\Delta\tau).
\end{equation}

\item \textit{Space-space} component:

\begin{align}
    N_{\bk}^{ij}(\tau,\tau')
    &=\frac{\pi \mathfrak{q}^2}{3}\left[(k^ik^j-k^2\delta^{ij})+\delta^{ij}\partial_\tau\partial_{\tau'}\right] \, \mathcal I_k(\tau-\tau').
\end{align}
\end{enumerate}

\noindent Collecting the components, the kernel can be represented by the differential operator
\begin{equation}
\label{eq:N_matrix_operator}
    N_{\bk}^{\mu\nu}(\tau,\tau')
    \doteq \frac{\pi \mathfrak{q}^2}{3}
    \begin{pmatrix}
        k^2
        &
        -i k^j\,\overrightarrow{\partial}_{\tau'}
        \\[6pt]
        i k^i\overleftarrow{\partial}_{\tau}\,
        &
        (k^ik^j-k^2\delta^{ij})
        +
        \delta^{ij}
        \overleftarrow{\partial}_{\tau}\,
        \overrightarrow{\partial}_{\tau'}
    \end{pmatrix} \, \mathcal I_k(\tau-\tau'),
\end{equation}
where $\doteq$ denotes equality up to boundary terms after integration by parts. and the arrows indicate which photon field is differentiated. Substituting \cref{eq:N_matrix_operator} into the photon contraction gives
\begin{align}
    N_{\bk}^{\mu\nu}(\tau,\tau') \left\langle A_\mu(\tau,\bk) A_\nu(\tau',-\bk)\right\rangle_{\mathrm{st}}
    &\doteq
    \frac{ \pi \mathfrak{q}^2}{3}\,
    \left\langle
    \Big[ k^2A_0(\tau,\bk)A_0(\tau',-\bk)\right.-i k^i A_0(\tau,\bk)\partial_{\tau'}A_i(\tau',-\bk)\nonumber\\
    &\quad +i\,\partial_\tau A_i(\tau,\bk)\,k^i A_0(\tau',-\bk)+\partial_\tau A_i(\tau,\bk) \partial_{\tau'}A_i(\tau',-\bk)\nonumber\\
    &\quad \left.-(k^2\delta^{ij}-k^ik^j)A_i(\tau,\bk)A_j(\tau',-\bk)\Big]
    \right\rangle_{\mathrm{st}}\times \mathcal I_k(\tau-\tau').
\label{eq:noise_AA_explicit}
\end{align}
To express this result in a manifestly gauge-invariant form, we define
\begin{align}
    &F_{\mu\nu}(\tau,\bk)\equiv\left\{
    \begin{aligned}
        &F_{0i}(\tau,\bk)=\partial_\tau A_i(\tau,\bk)-i k_iA_0(\tau,\bk)\\
        &F_{ij}(\tau,\bk)=i\left[k_iA_j(\tau,\bk)-k_jA_i(\tau,\bk)\right],
    \end{aligned}
    \right.
    \end{align}
and contracting both tensors, we obtain
\begin{equation}
    -\frac{1}{2}F_{\mu\nu}(\tau,\bk)F^{\mu\nu}(\tau',-\bk)=F_{0i}(\tau,\bk)F_{0i}(\tau',-\bk)-\frac{1}{2}F_{ij}(\tau,\bk)F_{ij}(\tau',-\bk).
\end{equation}
Expanding the right-hand side reproduces exactly the expression in
square brackets in \cref{eq:noise_AA_explicit}. Therefore,
\begin{equation}
    N_{\bk}^{\mu\nu}(\tau,\tau')\left\langle A_\mu(\tau,\bk)A_\nu(\tau',-\bk)\right\rangle_{\mathrm{st}}\doteq\frac{\pi \mathfrak{q}^2}{3}\,\mathcal I_k(\tau-\tau')\left\langle - \tfrac12 F_{\mu\nu}(\tau,\bk)F^{\mu\nu}(\tau',-\bk)\right\rangle_{\mathrm{st}}.
\label{eq:noise_FF}
\end{equation}
For completeness, we also compute the causal kernel 
\begin{equation}
    \mathcal J_k(\Delta\tau)\equiv\int_k^\infty d\omega\,\left(e^{-i\omega\Delta\tau} -e^{i\omega\Delta\tau}\right)
    =
    i\,\frac{2\cos(k\Delta\tau)}{\Delta\tau}.
\end{equation}
Following the same steps, we obtain
\begin{equation}
    \Pi_{\bk}^{\mu\nu}(\tau,\tau')\left\langle A_\mu(\tau,\bk)A_\nu(\tau',-\bk)\right\rangle_{\mathrm{st}}\doteq\frac{2\pi \mathfrak{q}^2}{3}\,\Theta(\tau-\tau')\mathcal J_k(\tau-\tau')\left\langle - \tfrac12 F_{\mu\nu}(\tau,\bk)F^{\mu\nu}(\tau',-\bk)\right\rangle_{\mathrm{st}}.
\end{equation}

\section{Flat-space limit and the four-momentum representation}
\label{sec:frequency_resolved_limit}
Our stochastic result is naturally formulated in the mixed representation $F_{\mu\nu}(\tau,\bk)$, which allows for a general time-dependent background without assuming a well-defined frequency. As a consistency check, one may take the flat-space limit, $a(\tau)\to1$, and restrict to stationary backgrounds admitting a well-defined frequency representation, in which case the mixed representation reduces to the four-momentum formulation $F_{\mu\nu}(\omega,\bk)$.

When the background admits a temporal Fourier representation,
\begin{equation}
    F_{\mu\nu}(\tau,\bk)
    =
    \int d\omega\  F_{\mu\nu}(\omega,\bk)\, e^{-i\omega\tau},
\end{equation}
with $K^\mu=(\omega,\bk)$, the mixed representation can be related directly to the usual four-momentum description. The memory kernel of \cref{eq:Ikernel} is in this convention the indicator function of the timelike region,
\begin{equation}
    \mathcal I_k(\tau-\tau')
    =
    \int d\omega  \ \Theta(\omega^2-k^2)\, e^{-i\omega(\tau-\tau')},
\label{eq:Ik_frequency} 
\end{equation}
so that $\widetilde{\mathcal I}_k(\omega)=\Theta(-K^2)$. Performing the two time integrations, each of which supplies a factor of $2\pi$, one finds
\begin{equation}
    \int d\tau\, d\tau'\ \mathcal I_k(\tau-\tau')\, F_{\mu\nu}(\tau,\bk)\, F^{\mu\nu}(\tau',-\bk)
    =
    (2\pi)^2 \int d\omega\ \Theta(-K^2)\, F_{\mu\nu}(\omega,\bk)\, F^{\mu\nu}(-\omega,-\bk).
\label{eq:mixed_to_frequency}
\end{equation}
Thus, whenever the time dependence is sufficiently long lived that a well-resolved frequency $\omega$ can be assigned, the stochastic expression reduces to the standard four-momentum form \cite{VicenteGarcia-Consuegra:2025lkh}
\begin{equation}
    2\,{\rm Im}\Gamma_{\rm eff}
    =
    \frac{2\pi^3\mathfrak q^2}{3} \int d^4K\  \Theta(-K^2)\,
    \left\langle -F_{\mu\nu}(K)F^{\mu\nu}(-K) \right\rangle_{\rm st}+\mathcal{O}(\mathfrak{q}^3).
\label{eq:ImGamma_fourmomentum}
\end{equation}
The mixed representation is therefore the more general description. It remains applicable to finite-duration or non-stationary stochastic backgrounds for which no sharply defined frequency exists, while \cref{eq:ImGamma_fourmomentum} is recovered in the frequency-resolved limit.

\phantomsection

\addcontentsline{toc}{section}{References}

%\bibliography{bibliography_m0}

\providecommand{\href}[2]{#2}\begingroup\raggedright\endgroup

\end{document}